\documentclass[
aip,
jcp,
floatfix,
 amsmath,amssymb,
preprint,%
 reprint,%
]{revtex4-1}

\usepackage{graphicx}
\usepackage{dcolumn}
\usepackage{bm}

\usepackage[utf8]{inputenc}
\usepackage[T1]{fontenc}
\usepackage{mathptmx}
\usepackage{etoolbox}
\usepackage{xcolor}

\makeatletter
\def\@email#1#2{%
 \endgroup
 \patchcmd{\titleblock@produce}
  {\frontmatter@RRAPformat}
  {\frontmatter@RRAPformat{\produce@RRAP{*#1\href{mailto:#2}{#2}}}\frontmatter@RRAPformat}
  {}{}
}%
\makeatother

\begin{document}

\preprint{AIP/123-QED}

\title{Structural and dynamic anomalous properties of TIP4P/2005 water at extreme pressures}
\author{Jos\'e Mart\'in-Roca}
 \affiliation{Departamento de Estructura de la Materia, F\'isica T\'ermica y Electr\'onica, Universidad Complutense de Madrid, Madrid, Spain}
\author{Alberto Zaragoza}%
\affiliation{Departamento de Matem\'aticas y Ciencias de Datos, Universidad San Pablo-CEU, CEU Universities, Madrid, Spain
}%
\author{Frédéric Caupin}
 \affiliation{Institut Lumi\`ere Mati\`ere, Universit\'e Claude Bernard Lyon 1, CNRS, Institut Universitaire de France, F-69622, Villeurbanne, France}
\author{Chantal Valeriani}
 \affiliation{Departamento de Estructura de la Materia, F\'sica T\'ermica y Electr\'onica, Universidad Complutense de Madrid, Madrid, Spain}

\date{\today}

\begin{abstract}
Water shows numerous thermodynamic, dynamic, and structural anomalies. Recent experiments [Eichler \textit{et al.} \textit{Phys. Rev. Lett.} \textbf{134}, 134101 (2025)], based on measurements of shear and bulk viscosities of liquid water up to 1.6 GPa, have reported the existence of a minimum in the variation of the structural relaxation time $\tau_\alpha$ with pressure at room temperature. Here we investigate this and related properties with molecular dynamics simulations of the TIP4P/2005 water model, performed at extreme pressures commensurate with the experiments. Specifically, we compute  dynamic (self-diffusion, shear and bulk viscosities, and structural relaxation time) and structural (oxygen-oxygen radial distribution function and structure factor, translational order parameter) properties {\color{black} down to 220 K and} up to 2.7 GPa. We find good agreement with the experimental observations, and confirm the existence of a minimum in $\tau_\alpha$. The microscopic information obtained from the simulations suggests that this anomaly is connected with the sudden reorganization of the hydrogen bond network induced by pressurization.
\end{abstract}

\maketitle

\section{\label{sec:Introduction}Introduction}

For decades, the water phase diagram has been a central topic in condensed matter physics. Not just for its interest as the most relevant compound in our lives, but also because of the intellectual challenge it has brought.\cite{gallo2016water} Its phase diagram presents around 20 solid phases,\cite{salzmann2019advances,loerting2020open}  while the liquid one exhibits a range of unusual properties such as density maxima at 4~ºC, high heat capacity or anomalous compressibility under certain conditions. These conditions may include high positive\cite{starr1999dynamics,brodholt1993simulations,singh2017pressure,mussa2023viscosity} and negative pressures\cite{gonzalez2016comprehensive,holten2017compressibility,biddle2017two}, confinement \cite{zangi2004water,zaragoza2019molecular,corti2021structure} or low temperatures at which liquid water is supercooled, metastable with respect to ice\cite{poole1992phase,dolan2007metastable,kim2017maxima,goy2018shrinking,goy2022refractive,coronas2024phase}. To explain the pronounced anomalies of water, an intriguing proposal has been put forward: water at low temperature may exhibit liquid polyamorphism, i.e. it may exist in two distinct liquid forms, separated by a first order transition terminating at a liquid-liquid critical point (LLCP) buried deep in the supercooled region. This scenario is supported by simulations\cite{poole1992phase,sciortino2025constraints}, extrapolations of equations of state\cite{holten2012entropy,caupin2019thermodynamics}, and experiments\cite{mishima1998decompression,kim2020experimental}. \textcolor{black}{Even at temperature above the LLCP, water can be described by two-state models as a non-ideal mixture of two interconverting local structures: an open, hydrogen-bonded, low-density and low-entropy state, favored at low pressure and temperature, and a collapsed, disordered, high-density and high-entropy state, favored at high pressure and temperature\cite{Anisimov_thermodynamics_2018,caupin2019thermodynamics}}

Already at ambient temperature, water exhibits striking dynamic anomalies. Usually, the dynamics in a liquid slows down with increasing pressure, in line with simple considerations based on the free volume theory\cite{doolittle1951studiesa,cohen1959molecular}. Water at moderate pressures shows the opposite behavior, with an increase in diffusivity\cite{prielmeier1988pressure,harris1997self}, and a decrease in viscosity\cite{bett1965effect,singh2017pressure}, before recovering a normal behavior at sufficiently high pressure. The structural relaxation time $\tau_\alpha$ was also reported to decrease with increasing pressure~\cite{bencivenga2009temperature}. Recently, experiments at higher pressure revealed the existence of a minimum in $\tau_\alpha$ around 0.5 GPa at room temperature\cite{eichler2025shear}. In \textcolor{black}{the two-state picture, $\tau_\alpha$ has been interpreted as a characteristic time for interconversion between the two local structures\cite{hall1948origin}}. In Ref.~\onlinecite{eichler2025shear}, it was speculated that the minimum in $\tau_\alpha$ could be related to a structural anomaly at which the interconversion rate would reach a maximum. \textcolor{black}{Indeed, simulations with various water models revealed a structural anomaly at high pressure: a line of minima in the translational order parameter $t$\cite{errington2001relationship,agarwal2011thermodynamic}. These studies revealed a hierarchy between anomalies: in the temperature-pressure plane, the line of $t$ anomalies circles the line of diffusivity anomaly, which itself surrounds the line of density anomaly.}

\textcolor{black}{Our motivation is three-fold. First, the key experimental quantity obtained in Ref.~\onlinecite{eichler2025shear} is the bulk viscosity of water, for which very few simulations are available; in this previous work, we reported simulated values of shear and bulk viscosity over a broad pressure range at 300 K. Here we extend the study to two lower isotherms, 260 and 280 K. Second, to our knowledge, the existence of a pressure minimum in $\tau_\alpha$ has been studied with molecular dynamics simulations only for the SPC/E potential\cite{starr1999dynamics}. SPC/E has been developed 38 years ago\cite{Berendsen_missing_1987} and played a major role in simulations of water. However, 20 years ago, the TIP4P/2005 model appeared\cite{abascal2005general}, and since has been recognized as the potential showing the best performance among rigid non-polarizable models\cite{Vega_simulating_2011}. We thus perform simulations with TIP4P/2005 to locate the line of minima in $\tau_\alpha$ and compare it to SPC/E and experiments. Third, the relation of the line of $\tau_\alpha$ minima to the nested pattern formed by other lines of anomalies\cite{errington2001relationship,agarwal2011thermodynamic} has not been discussed before. In the present work, we provide a complete picture of all anomalies for experiments and simulations with SPC/E and TIP4P/2005.}

Section~\ref{sec:simul} \textcolor{black}{gives} details about the simulations and their analysis. The results are reported in Section~\ref{sec:results}. \textcolor{black}{Self-diffusion, shear and bulk viscosities are obtained along three isotherms (260, 280, and 300 K), whereas data for oxygen-oxygen radial distribution function and structure factor, translational order parameter, and structural relaxation time cover 5 isotherms (220, 240, 260, 280, and 300 K). Section~\ref{sec:discussion} discusses the decoupling between dynamic quantities and compares the pattern formed by the lines of anomalies in experiments and simulations.}

\section{Simulation details\label{sec:simul}}

\subsection{Simulation parameters}

All molecular dynamics (MD) simulations were performed using the LAMMPS simulation package \cite{thompson2022lammps}. Water molecules were modeled using the rigid, non-polarizable TIP4P/2005 force field \cite{abascal2005general}, which provides an accurate description of the phase diagram and thermodynamic properties of water. The geometry of the water molecules was constrained using the SHAKE algorithm, allowing for an integration time step of $2\,\mathrm{fs}$. To investigate the behavior of water across a wide range of states, simulations were conducted {\color{black} along five isotherms: 220 K, 240 K,} 260 K, 280 K, and 300 K. For each temperature, a series of densities ranging from 0.910 g cm$^{-3}$ to 1.35 g cm$^{-3}$ was explored. The size of the cubic simulation box was adjusted for each target density. To prevent spontaneous crystallization when the liquid becomes metastable at high pressure, all simulations contained a system size of $N = 216$ water molecules, following the established practice of Montero de Hijes et al. \cite{montero2018viscosity}. For each state point ($T$, $\rho$), a minimum of 20 independent simulation runs were initiated from different initial configurations and velocity distributions to ensure robust statistical sampling. Because the calculation of bulk viscosity is more conveniently performed in the microcanonical ensemble~\cite{bertolini1995stress,fanourgakis2012determining,jaeger2018bulk,hafner2022sampling}, the simulation protocol for each run consisted of two stages: (a) \textbf{Equilibration:} The system was first equilibrated in the canonical (NVT) ensemble for 20 ps. Temperature was maintained using the Nosé-Hoover thermostat with a relaxation time of 0.1 ps. (b) \textbf{Production:} Following equilibration, a 10 ns microcanonical (NVE) production run was performed to collect trajectory data for the analysis of dynamic and transport properties. {\color{black}Configurations were saved every 1 ps for subsequent analysis, and the relevant properties, such as the correlations used in the calculation of viscosities, were computed on the fly every 0.1 ps using the ave/correlate/long routine available in LAMMPS.}

\subsection{Analysis Tools\label{sec:tools}}

The saved trajectories were subsequently analyzed to extract the thermodynamic, structural, and dynamic properties of the system. The equation of state, namely the relationship between pressure ($P$), density ($\rho$), and temperature ($T$), was determined by calculating the average of the virial and kinetic contributions to the pressure tensor throughout the production phase of the simulation.

The transport coefficients, namely the shear viscosity ($\eta$) and bulk viscosity ($\kappa$), were calculated using the Green-Kubo formalism~\cite{bertolini1995stress,fanourgakis2012determining,jaeger2018bulk,hafner2022sampling}. {\color{black} This method relates the viscosities to the time integral of the auto-correlation functions of the stress tensor elements $\sigma_{\alpha\beta}=-p_{\alpha \beta}$, where $p_{\alpha \beta}$ are the pressure tensor elements. The equilibrium pressure is given by $\langle  P\rangle =\frac{1}{3} \text{Tr}(\mathbf{P}(t))$, where $\text{Tr}(\mathbf{P}(t))$ is the time average of the trace of the pressure tensor $\mathbf{P}$}. The shear viscosity is given by the integral of the correlation of the off-diagonal elements of the stress tensor:
\begin{equation}
    \eta = \frac{V}{k_B T} \int_0^\infty \sum_{\alpha < \beta} \langle \sigma_{\alpha\beta}(t) \, \sigma_{\alpha \beta }(0) \rangle  dt, \quad  \alpha, \beta \in \{x, y, z\}
\end{equation}
the brackets $\langle \cdots \rangle$ represent an ensemble average, $V$ is the volume of the simulation box, $k_B$ is Boltzmann's constant, and $T$ the temperature. The bulk viscosity is obtained from the auto-correlation of the fluctuations in the trace of the pressure tensor:
\begin{equation}
    \kappa= \frac{V}{k_B T} \int_0^\infty  \left\langle \delta \Pi(t) \, \delta \Pi(0) \right\rangle  dt, 
\end{equation}
where $\delta \Pi(t) = \frac{1}{3} \text{Tr}(\mathbf{P}(t)) - P$ is the instantaneous fluctuation of pressure.

The self-diffusion coefficient ($D_{PBC}$) in the box with periodic boundary conditions was obtained by analyzing the mean squared displacement (MSD) of the oxygen atoms and applying the Einstein relation
\begin{equation}
    D_{PBC} = \lim_{t \to \infty} \frac{1}{6t} \langle | \mathbf{r}_i(t) - \mathbf{r}_i(0) |^2 \rangle
    \label{eq:Dcorr}
\end{equation}
$D_{PBC}$ was then corrected for finite-size effects using the correction \cite{montero2018viscosity}
\begin{equation}
D = D_{PBC} + 2.837\frac{k_{B}T}{6 \pi \eta L_{\text{Box}}}
\label{eq:diffus2}
\end{equation}
where $k_B$ is Boltzmann's constant, $T$ is the temperature, 
$\eta$ is the shear viscosity, and $L_{\text{Box}}$ is the length of the simulation box. This correction is essential for small systems like ours ($N=216$) to account for the hydrodynamic interactions that are truncated by periodic boundary conditions \cite{montero2018viscosity}.

Structural information was obtained by computing the radial distribution functions (RDFs) between oxygen-oxygen $g(r)$ and the oxygen-oxygen structure factor $S(q)$, defined as
\begin{equation}
S(q) = \left \langle  \frac{1}{N} \sum_{j=1}^N \sum_{k=1}^N \exp\left[-i \, q \, \hat{n}  \cdot  (\vec{r}_j-\vec{r}_k) \right] \right\rangle_{\hat{n}}
\label{sq_eq}
\end{equation}
where \textit{$\vec{r}_m$} is the position of molecule \textit{m}, \textit{q} is the scattering vector and $\hat{n}$ represents the unit vector defining the scattering vector direction. In this case, the brackets  $\langle \cdots \rangle_{\hat{n}}$ represent an average over different possible spacial directions. 

We perform coordination analysis using the definition of the coordination number 
\begin{equation}
    n(r_c) = 4\pi \rho \, \int_0^{r_c} r^2 \, g(r) \, dr,
    \label{eq:coordination}
\end{equation}
computed until a given cut-off distance, $r_c$. 

Following Refs.~\onlinecite{errington2001relationship,agarwal2011thermodynamic},  
we measure the extent of pair correlations present between oxygen atoms in molecules (tetrahedral centers) computing the pair correlation or
translational order parameter, $t$, defined as
\begin{equation}
    t= \frac{1}{\xi_c} \, \int_0^{\xi_c} \left| g(\xi) -1\right| \, d\xi 
    \label{eq:t}
\end{equation}
where $\xi = r \, \rho^{1/3}$; $r$ is the pair separation; $g(\xi)$ is the radial distribution function; and $\xi_c$ is a suitably chosen cutoff distance that we choose as the half of the box $L/2$, since it performs a plateau value in $t$ for all temperatures and pressures (at long enough distances).

Additionally, the relaxation dynamics was probed by calculating the self-intermediate scattering function
\begin{equation}
F(Q,t) = \left \langle  \frac{1}{N} \sum_{j=1}^N \sum_{k=1}^N \exp\left\{-i \, q \, \hat{n}  \cdot  \left[\vec{r}_j(t)-\vec{r}_j(0)\right] \right\} \right\rangle_{\hat{n}}
\label{eq:F}
\end{equation}
where $Q$ is the value of the wavevector for which $S(q)$ is maximum. The decay of $F(Q, t)$ was analyzed by fitting it to a stretched double-exponential function\cite{sciortino1996supercooled,gallo1996slow,de2016mode}
\begin{equation}
    F(Q,t) = \left[1-A(Q)\right] \, \exp \left[-\left( \frac{t}{\tau_{s}} \right) ^{2}\right] + A(Q) \, \exp \left[-\left( \frac{t}{\tau_{\alpha}} \right) ^{\beta}\right]
\label{eq:exp_fit}
\end{equation}
This fit allows for the characterization of relaxation timescales as a function of wavevector $Q$, pressure, and temperature. In this expression, $\tau_{s}$ represents the rapid short-time relaxation and $\tau$$_{\alpha}$ indicates the $\alpha$ relaxation time. The Kohlrausch exponent, $\beta$, describes the stretched exponential decay observed at longer times. The plateau height $A(Q)$, also known as the  Lamb-M\"osbauer factor \cite{gotze2009complex}, is a direct measure of the localization strength or the confinement of a molecule within the cage formed by its nearest neighbors. To estimate the characteristic cage size, we employed the Gaussian approximation. Within this framework, $A(Q)$ is related to the mean squared displacement of a molecule within its cage by
\begin{equation}
    A(Q) = \exp \left[ 
- \frac{1}{3} \, r_\mathrm{cage}^2 \, Q^2 \right]
\label{eq:cage_rad}
\end{equation}
where $r_\mathrm{cage}$ is the average effective cage radius.

\section{Results\label{sec:results}}

\subsection{Equation of state}

\begin{figure}[tt]
    \centering
{\hspace{-0.5cm}\includegraphics[width=1.05\linewidth]{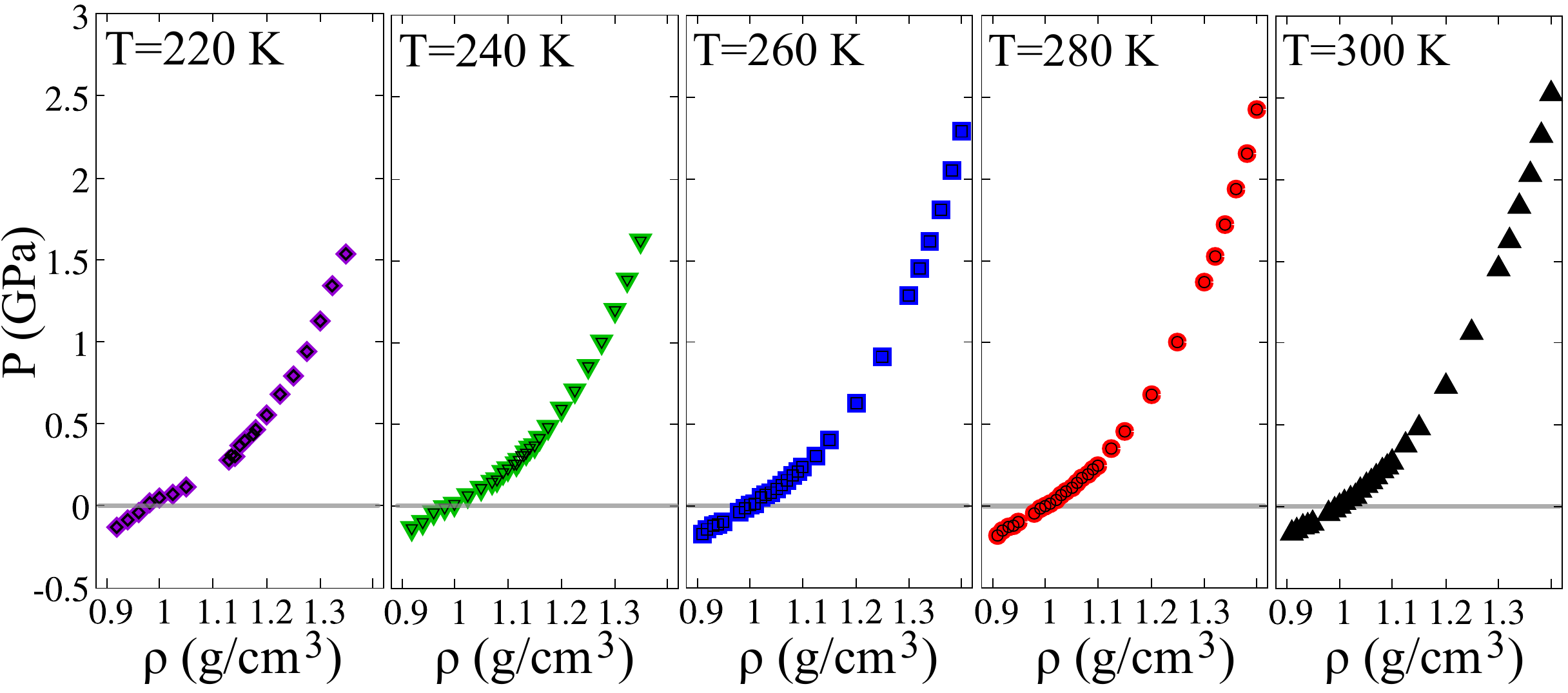}}
    \caption{Equation of state calculated for T = \textcolor{black}{ 220 K (a, purple diamonds), 240 K (b, green inverted triangles), 260 K (c, blue squares), 280 K (d, red circles), and 300 K (e, black triangles).}
    }
    \label{fig:press_vs_dens}
\end{figure}
Figure \ref{fig:press_vs_dens} shows the variation of pressure $P$ as a function of density $\rho$ along the \textcolor{black}{five} isotherms studied: \textcolor{black}{220, 240,} 260, 280, and 300 K. The observed behavior for the three highest temperatures is very similar, due to the small variation of compressibility in this range. \textcolor{black}{At the lowest temperature, compressibility increases (another anomaly of water), and, for the same density interval, the pressure range is reduced.} In the following, we report all results as a function of $P$ instead of $\rho$, to allow an easier comparison with experiments.

\subsection{Viscosities\label{sec:visc}}

\begin{figure}[b]
    \centering
    \includegraphics[width=0.9\linewidth]{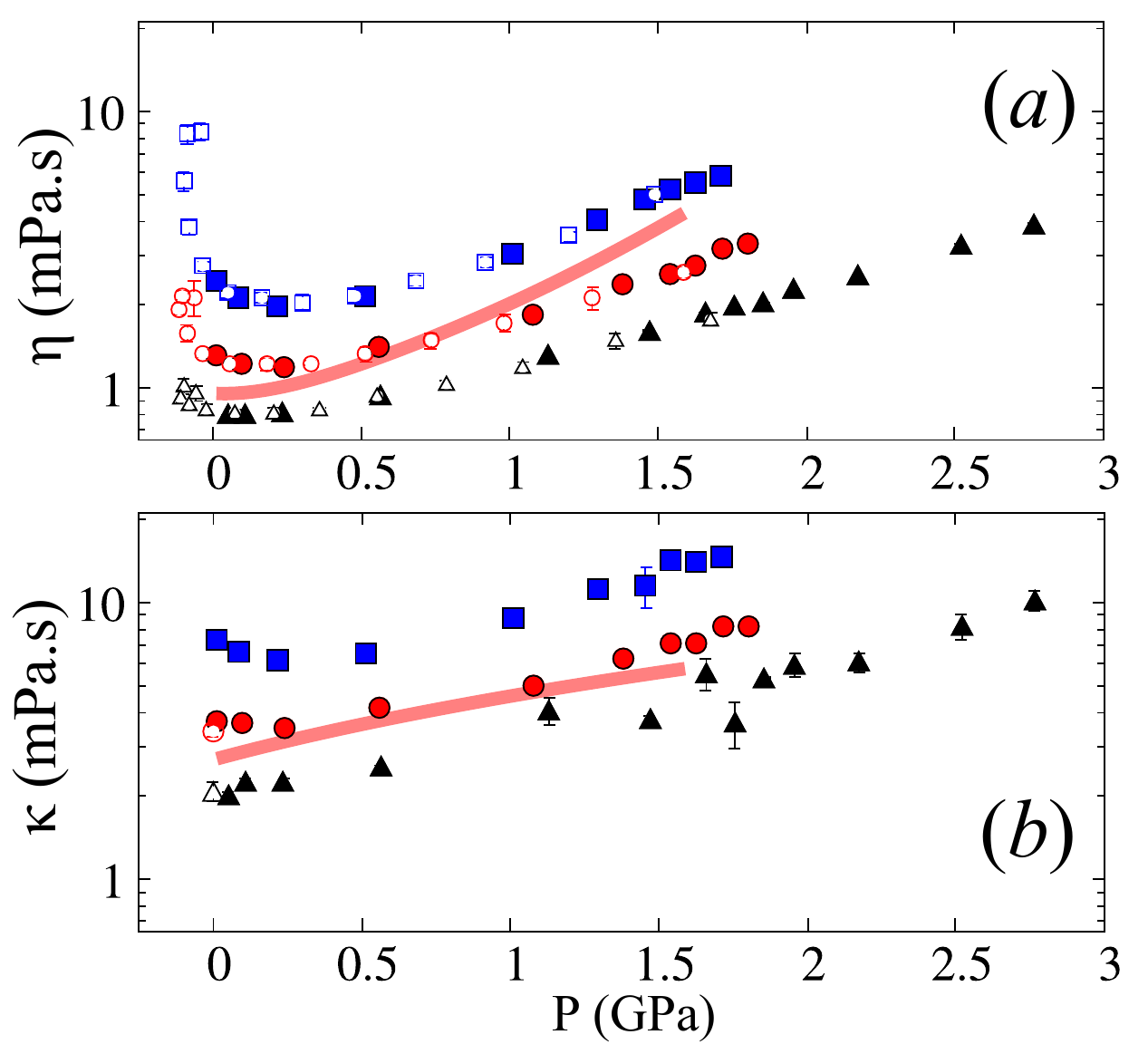}
    \caption{Shear $\eta$ (a) and bulk $\kappa$ (b) viscosities as a function of pressure $P$ computed for $T =$ 260 K (blue squares), 280 K (red circles), and 300 K (black triangles). Filled symbols correspond to the present work, and empty symbols to previous simulations from Ref.~\onlinecite{montero2018viscosity} (a) and Ref.~\onlinecite{fanourgakis2012determining} (b). The solid red curves display \textcolor{black}{polynomial fits to} the experimental results at 295 K~\cite{eichler2025shear} \textcolor{black}{(note that the experimental minima in $\eta$ and $\kappa$ are not well captured by the fits on this scale)}.}
    \label{fig:visco_vs_press}
\end{figure}

 Figure \ref{fig:visco_vs_press} displays shear (a) and bulk (b) viscosity as a function of pressure. \textcolor{black}{All values are given in Table~\ref{table: Diffusion}.}
 
 In both cases, cooling at constant pressure increases viscosity. \textcolor{black}{In experiments below 306 K, $\eta$ along an isotherm goes through a minimum, which becomes more pronounced at lower temperature, while its position shifts to higher pressures\cite{bett1965effect,singh2017pressure}. This trend was already correctly captured} in our previous simulations at lower pressures\cite{montero2018viscosity}. Our present work is in excellent agreement with these previous results, and reaches \textcolor{black}{higher pressures}, confirming that water recovers a normal behavior, with a steady increase of shear viscosity at high pressure. The experimental variation at 295 K (solid red curve) is better reproduced by the simulations at 280 K. 

Simulations of bulk viscosity in water are scarce, and limited to low pressures~\cite{bertolini1995stress,fanourgakis2012determining,jaeger2018bulk,hafner2022sampling}. Our present results (\textcolor{black}{Fig. \ref{fig:visco_vs_press}}b) agree with previous simulations, and cover a much broader pressure range, revealing a behavior of bulk viscosity 
($\kappa$) similar to that of shear viscosity: $\kappa$ along an isotherm goes through a minimum, which becomes more pronounced at lower temperature, while its position shifts to higher pressures. Once again, the experimental variation at 295 K (solid red curve) is better reproduced by the simulations at 280 K.

\begin{figure}[t]
    \centering
    \includegraphics[width=0.85\linewidth]{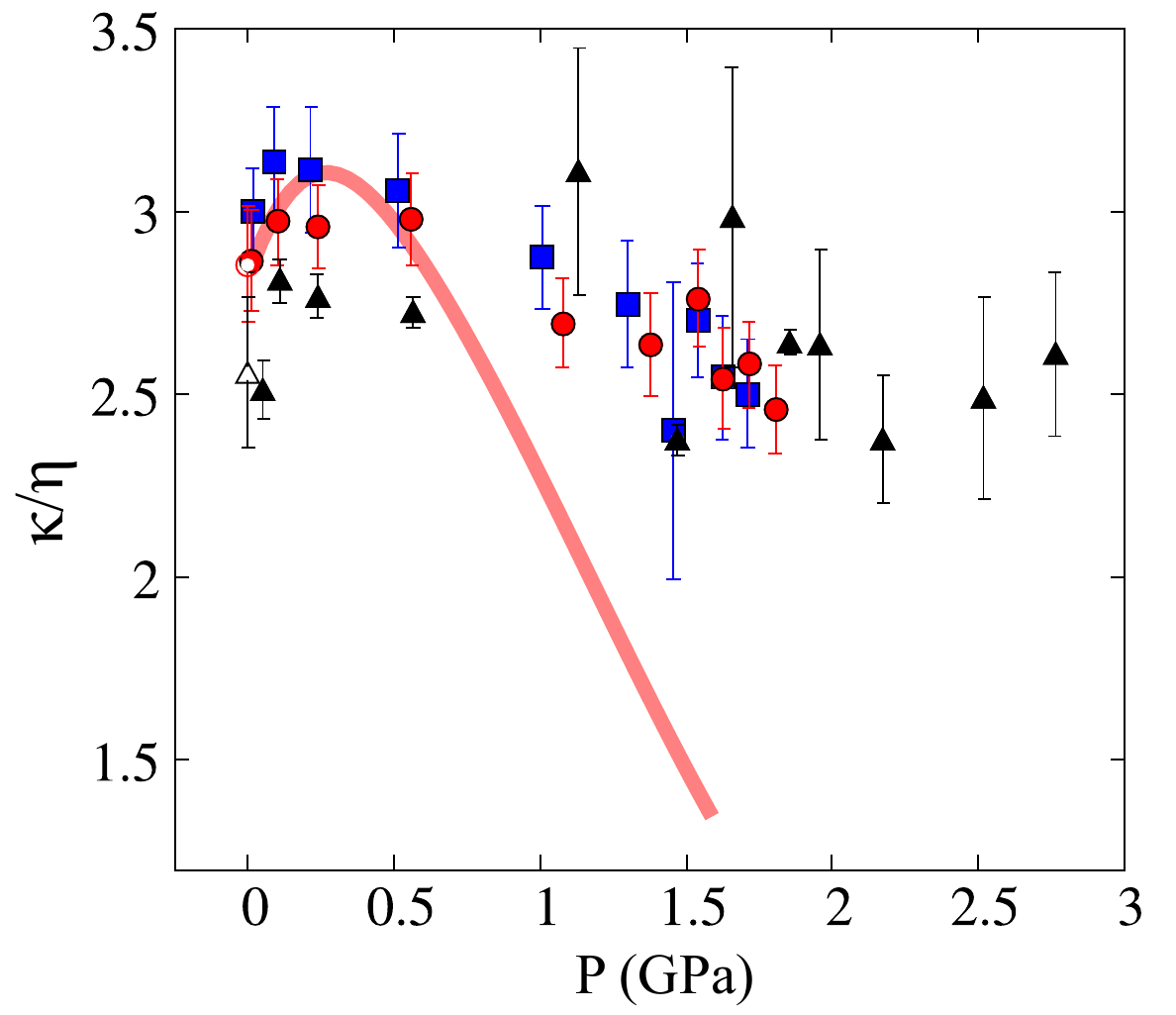}
  \caption{Ratio between bulk ($\kappa$) and shear viscosity ($\eta$) as a function of pressure $P$ for 260 K (blue squares) , 280 K (red circles) and 300 K (black triangles). The solid red curve displays the experimental results at 295 K~\cite{eichler2025shear}.
   }
  \label{fig:visco_ratio}
\end{figure}

After analyzing the bulk and shear viscosities separately, we calculated their ratio ($\kappa/\eta$).
Our results, shown in figure \ref{fig:visco_ratio}, indicate a mild maximum of ($\kappa/\eta$)
at low pressure, followed by a decrease, which seems slightly more pronounced at lower temperature. The same behavior is observed in experiments (solid red curve), albeit with a steeper decrease at high pressure.


These results suggest that at low temperatures, water's dynamics are highly sensitive to pressure, likely due to its proximity to hypothesized thermodynamic anomalies and glass transition phenomena \cite{gallo2016water}. \textcolor{black}{The two viscosities of water recover a normal behavior at high temperature and pressure.}


\begin{figure}[t]
    \centering
    \includegraphics[width=0.85\linewidth]{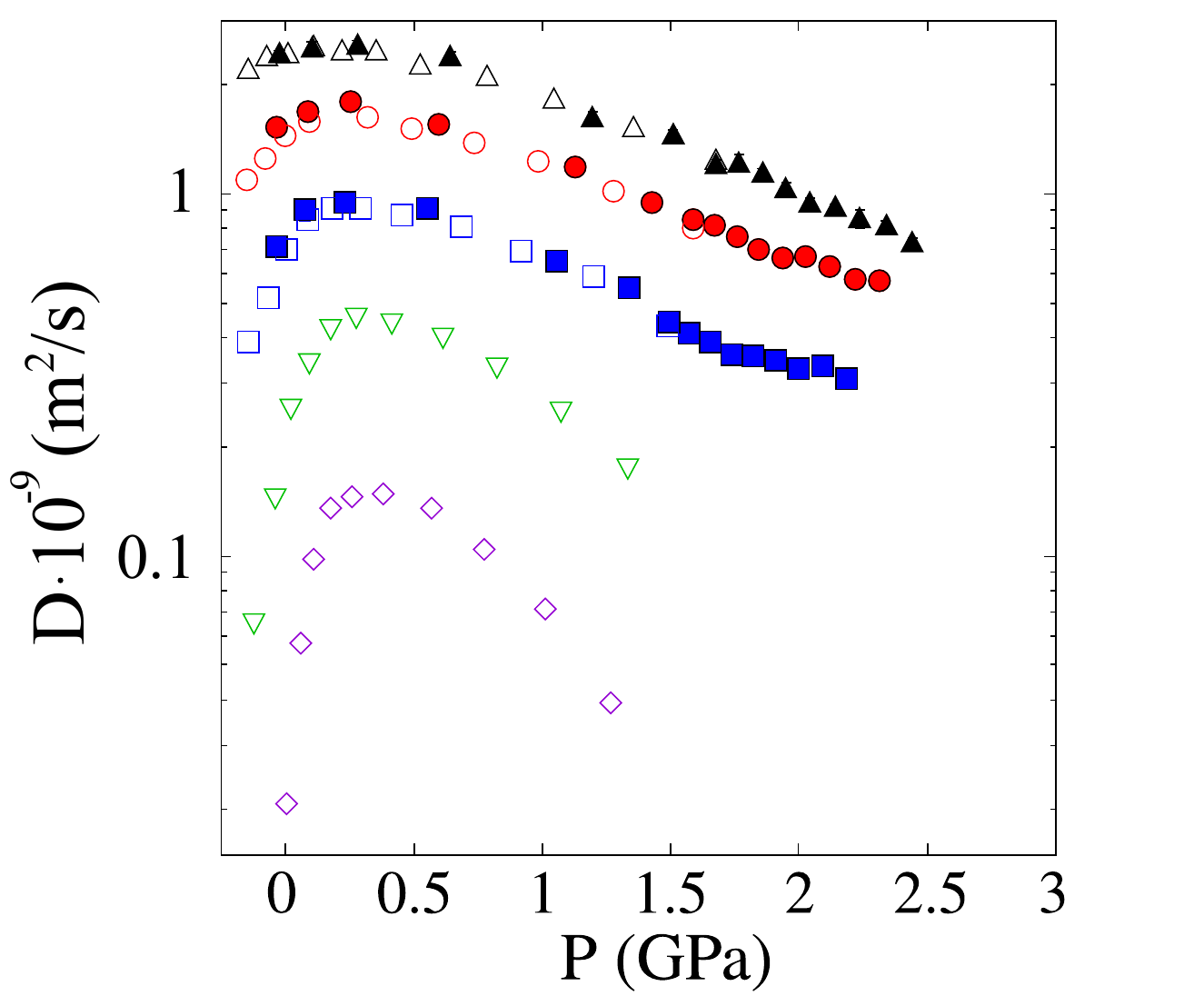}
    \caption{Self-diffusion coefficient $D$ as a function of pressure $P$ for {\color{black} 220~K (purple diamonds), 240~K (green inverted triangles)}, 260 K (blue squares) , 280 K (red circles) and 300 K (black triangles).  Filled symbols correspond to the present work, and empty symbols to previous simulations from Ref.~\onlinecite{montero2018viscosity}
    }
    \label{fig:diff_press}
\end{figure}

\subsection{Self-diffusion coefficient\label{sec:D}}

Figure \ref{fig:diff_press} displays the self-diffusion coefficient $D$ as a function of pressure along the three isotherms studied \textcolor{black}{in the present work (see Table~\ref{table: Diffusion} for $D$ values and parameters used for the correction with Eq.~\ref{eq:Dcorr}), together with our previous results down to lower temperatures\cite{montero2018viscosity}.} 

\begin{figure*}[!]
    \centering
    \includegraphics[width=0.9\linewidth]{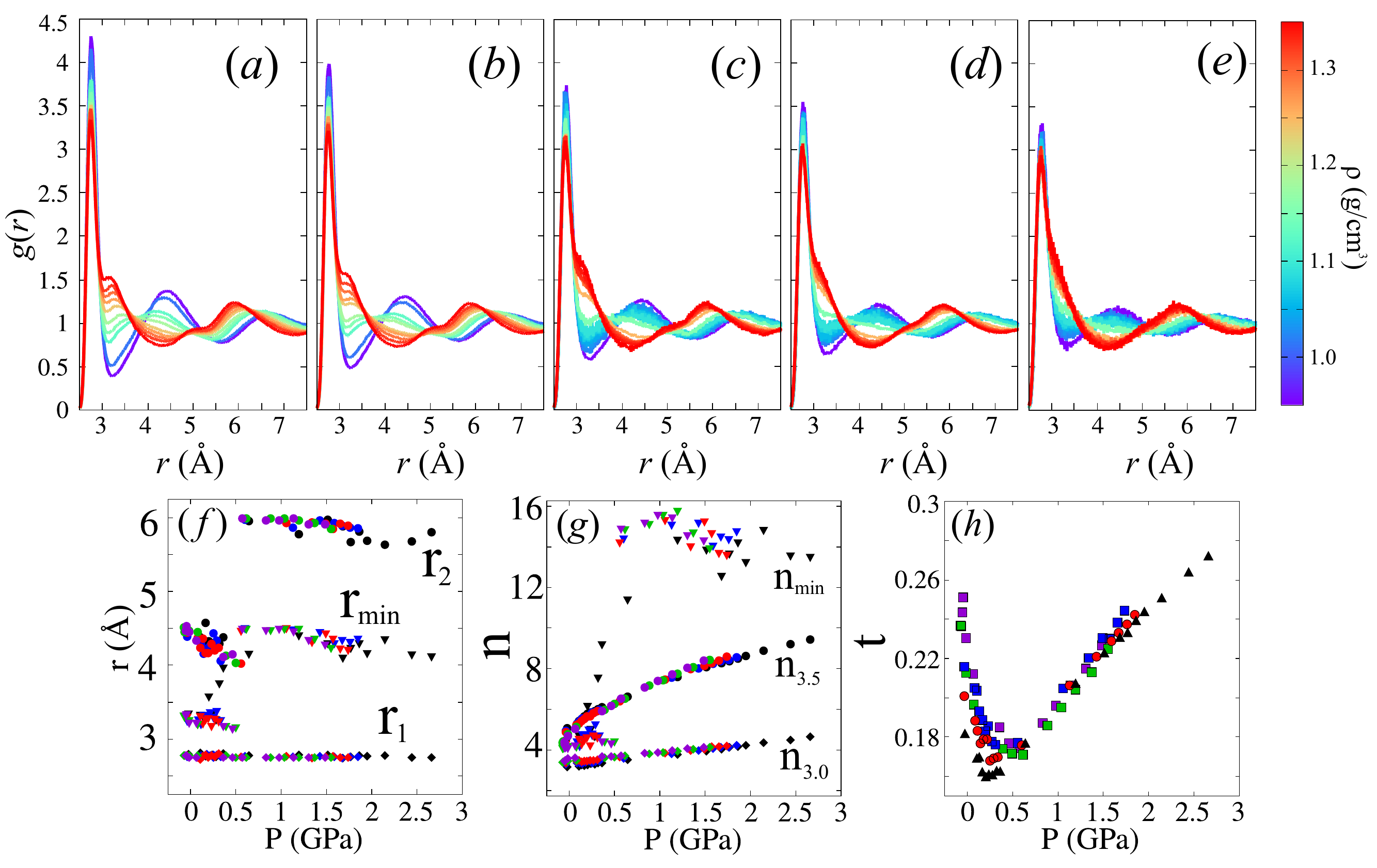}
    \caption{(a-e) Radial distribution function $g(r)$; (f) distance for the first minima in radial distribution function, $r_\mathrm{min}$ (inverted triangles), and distances for the two first maximum values, $r_1$ (diamonds) and $r_2$ (circles); (g) coordination number computed for the first minima in radial distribution function,  $n_\mathrm{min}=n(r_\mathrm{min})$ (inverted triangles), and for some constant distance $n_{3.0}=n(r_c=3.0$\AA$)$ (diamonds) and $n_{3.5}=n(r_c=3.5$\AA$)$ (circles); and (h) translational order parameter $t$, calculated for all densities/pressures at  temperatures $T =$ {\color{black} 220 K (purple), 240 K (green)}, 260 K (blue), 280 K (red), and 300 K (black).}
    \label{fig:RDF}
\end{figure*}

Cooling at constant pressure decreases diffusivity. \textcolor{black}{In experiments at room temperature and below, $D$ along an isotherm goes through a maximum, which becomes more pronounced at lower temperature, while its position shifts to higher pressures\cite{prielmeier1988pressure,harris1997self}. This trend was already correctly captured} in previous TIP4P/2005 simulations at lower pressures\cite{agarwal2011thermodynamic,montero2018viscosity}; for a direct comparison with experiments, see Fig. 2 of Ref.~\onlinecite{montero2018viscosity}. \textcolor{black}{Our present work is in excellent agreement with these previous results, and reaches higher pressures, confirming} that water recovers a normal behavior, with a steady decrease of diffusivity at high pressure. 


\subsection{Structural properties}

\subsubsection{Radial distribution function\label{sec:g}}

The top row of Fig.~\ref{fig:RDF} shows our results for the oxygen-oxygen radial distribution function, $g(r)$, along the {\color{black} five} isotherms.
At a temperature of 300 K and the lowest density ($\rho \approx 0.997$ g cm$^{-3}$), $g(r)$ exhibits the characteristic profile of liquid water, with two clearly separated maxima at short distances. The first maximum, corresponding to the first coordination shell, is located at $r \approx 2.85$ Å. The subsequent minimum, which defines the spatial boundary of this shell, is found at $3.3$ Å. The second maximum is weaker and located around $4.45$ Å.

As the density increases, the second maximum broadens and shifts to lower distances, until above $0.75\,\mathrm{GPa}$ it becomes a shoulder on the right flank of the first peak. This behavior reproduces that inferred from neutron diffraction\cite{soper2000structures}. The first coordination shell remains tetrahedral with about four water molecules at all pressures, while pressure makes the second shell collapses on the first, implying strong distortions in the hydrogen bond network.

The effect of temperature is distinctly different from that of density. {\color{black} At the lowest condition of 220 K}, the RDF peaks are significantly more intense and sharper compared to those at 300 K. This enhancement in peak definition indicates a stronger and more well-defined local order, a consequence of reduced thermal motion that allows the hydrogen-bond network to become more rigid and pronounced. However, the positions of the maxima and minima remain largely invariant with temperature at a fixed density. This key observation suggests that temperature primarily modulates the amplitude and sharpness of structural correlations (i.e., the strength and definition of the hydrogen-bond network), while density primarily governs their spatial extent (i.e., the distances at which these correlations occur).


Figure 5{\color{black} (f-h)} shows the properties extracted from $g(r)$. Panel (f) presents for the five isotherms the location of the first extrema in $g(r)$ as a function of pressure: the two first maxima at $r_1<r_2$, and the first minimum at $r_\mathrm{min}$. $r_1$ remains constant at $\sim 2.75$ \AA~ at all conditions, whereas at the structural crossover, $r_2$ and $r_\mathrm{min}$ jump from 4.4 to 5.8 \AA~ and from 3.3 to 4.4 \AA, respectively. {\color{black} Figure~5 (g)} shows the coordination number $n$ around a water molecule, calculated in various spheres. Within a radius of 3 \AA, $n_3$ remains close to 4 at all conditions; this was noticed in previous works~\cite{kalinichev1999structure,yan2007structure,fanetti2014structure,imoto2015water,skinner2016structure,prasad2022hydrogen}, in which further analysis showed that the average number of hydrogen bonds remains almost unchanged during compression. By computing $n$ in a 3.5 \AA~ sphere, it was shown that other molecules progressively occupy interstitial positions between the four hydrogen bonded ones, slowly increasing the corresponding coordination number $n_{3.5}$. We confirm these observations here with a continuous increase from 5 to 9 across the pressure range investigated. The coordination number computed up to the first minimum shows a more abrupt behavior, reflecting the corresponding jump in $r_\mathrm{min}$: $n_\mathrm{min}$ sharply jumps between two plateaus around 5 and 14 at the structural crossover around 0.5 GPa.

\begin{figure*}[t]
    \centering
    \includegraphics[width=0.99\linewidth]{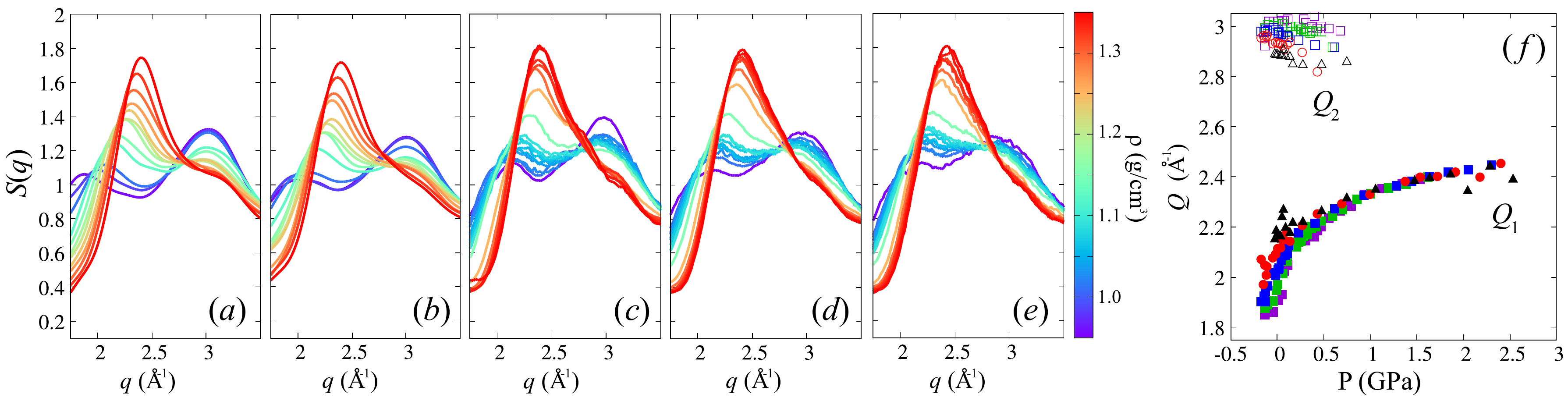}
    \caption{(a-e) Static structure factor for various densities at 220~K (a), 240~K (b), 260~K (c), 280~K (d) and 300~K (e). (f) Position of the two first maxima in $S(q)$, $Q_1$ (filled symbols) and $Q_2$ (empty symbols) as a function of pressure $P$, computed for $T =$ {\color{black} 220~K (purple), 240~K (green),}  260~K (blue squares), 280~K (red circles) and 300 K (black triangles).
    }
    \label{fig:Sq_vs_dens}
\end{figure*}

\subsubsection{Translational order parameter}

The translational order parameter $t$, defined in Eq.~\ref{eq:t}, measures the tendency of pairs of molecules to adopt preferential separations. {\color{black} Figure~\ref{fig:RDF} (h)} shows that this order parameter goes through a minimum on each isotherm, at a pressure which increases upon cooling. \textcolor{black}{Interestingly, this minimum is located close to the pressure at which $r_\mathrm{min}$ jumps (Fig.~\ref{fig:RDF}f).} As first pointed out by Errington and Debenedetti~\cite{errington2001relationship}, \textcolor{black}{the line of $t$ minima} defines a region of structural anomaly within which order decreases under compression (see Section~\ref{sec:discussion}).

\subsubsection{Structure Factor\label{sec:Sq}}

{\color{black}The panels (a-e)}  of Fig.~\ref{fig:Sq_vs_dens} show our results for the oxygen-oxygen static structure factor, $S(q)$, along the five isotherms.

$S(q)$ is related to the Fourier transform of $g(r)$, and, in line with the behavior of the latter, exhibits a clear structural crossover between low and high pressure. The low-density regime is characterized by the presence of two distinct maxima located at $Q \sim 2$ Å$^{-1}$ and $\sim 3$ Å$^{-1}$, a signature of the persistent tetrahedral ordering in the water network. In contrast, for densities $\rho \geq 1.2$ g·cm$^{-3}$, these two peaks merge into a single, broader maximum centered at $Q \sim 2.5$ Å$^{-1}$. This \textcolor{black}{qualitatively reproduces} the results of neutron diffraction (which directly access $S(q)$ rather than $g(r)$)\cite{soper2000structures}.
This merger signifies a breakdown of the tetrahedral hydrogen-bond network and a crossover to a more densely packed, distorted structure. {\color{black} This same density-induced crossover is observed at other temperatures}, underscoring that the change in packing is a fundamental response to pressure, albeit occurring at slightly different thresholds.

The effect of temperature is most apparent in the low-density regime. {\color{black} At lower temperatures (220 K - 260 K), the two peaks} at low $Q$ ($\leq 4$ Å$^{-1}$) \textcolor{black}{are} notably sharper and more intense, reflecting the enhanced intermediate-range order. Conversely, for high-density systems, the effect of temperature on $S(q)$ becomes negligible across the entire $Q$-range, indicating that the high-pressure environment dominates the structure, overwhelming the subtler effects of thermal energy.

To quantify this crossover, we precisely tracked in {\color{black} Fig. ~\ref{fig:Sq_vs_dens}.f.} the position of the two first maxima in $S(q)$, $Q_1 < Q_2$, across a finely spaced grid of pressures, including intermediate values not used for other properties. A clear discontinuity is observed, nearly independent of temperature. At 300 K, $Q_2$ remains at a nearly constant value of $\sim 2.9$ Å$^{-1}$ until pressure 0.9 GPa, where it disappears, because of the merging of the two peaks in $S(q)$. At 300 K, $Q_1$ increases from around $2\,$Å$^{-1}$ to 2.442 Å$^{-1}$ at the highest pressure explored. At {\color{black}the other temperatures explored}, the behavior is qualitatively identical. After pressure 0.9 GPa the values of $Q_1$ overlap at all three temperatures, confirming that the high-density structure is essentially temperature-independent in the investigated range.

\begin{figure}[b]
    \centering
    \includegraphics[width=0.9\linewidth]{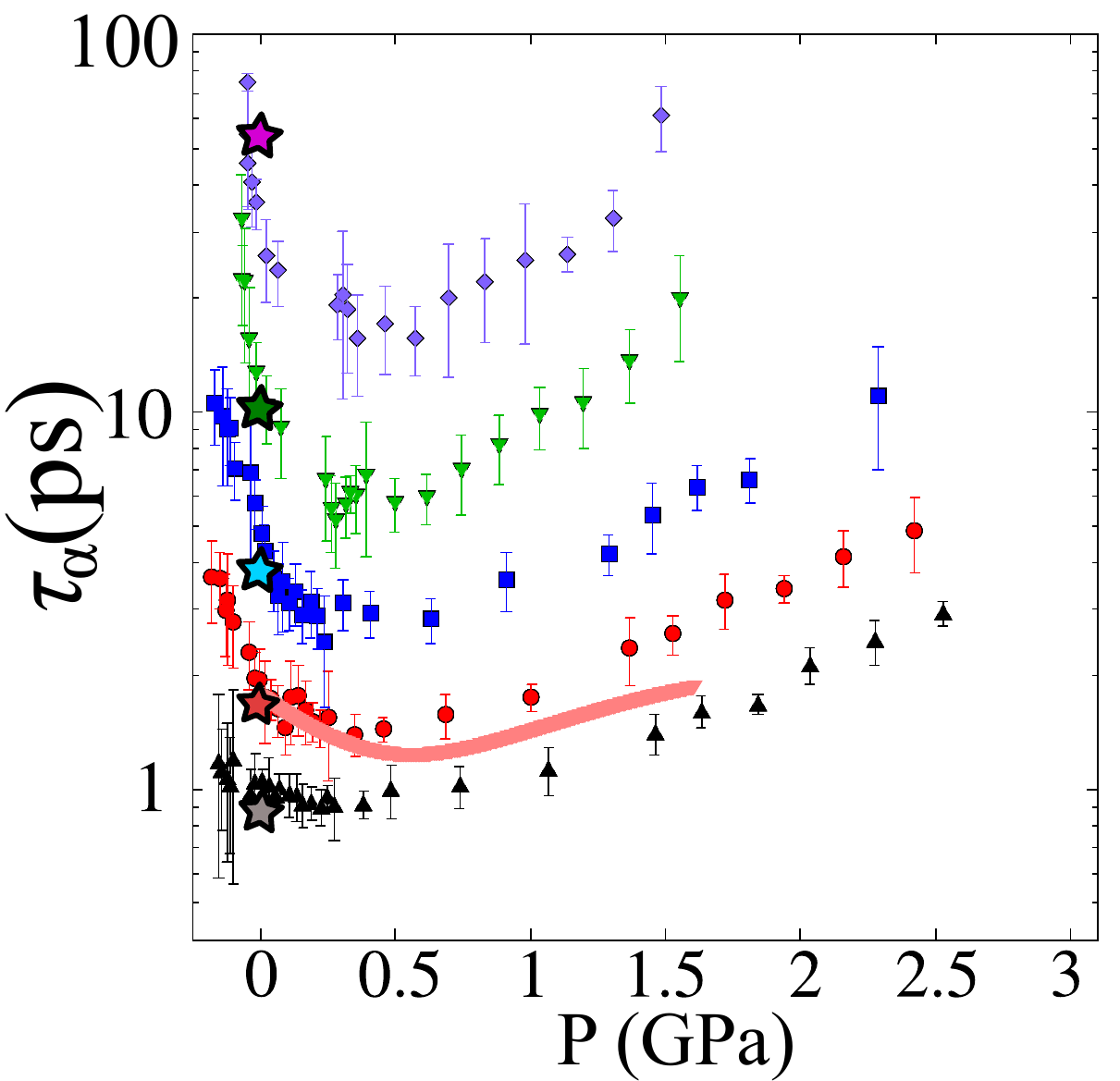}
    \caption{$\alpha$-relaxation time (obtained from fits of the self-intermediate scattering function $F(Q_1,t)$ with Eq.~\ref{eq:exp_fit}) as a function of pressure for {\color{black} 220~K (purple), 240~K (green),} 260~K (blue), 280~K (red), and 300~K (black). Stars symbols shown have been obtained at a fixed $Q=2.25$ Å$^{-1}$ in Ref.~\onlinecite{de2016mode}. The red solid curve displays the relaxation time obtained experimentally at 295 K~\cite{eichler2025shear}.}
    \label{fig:tau_alpha}
\end{figure}

\subsection{Structural relaxation}

\begin{figure}[t]
    \centering
    \includegraphics[width=1\linewidth]{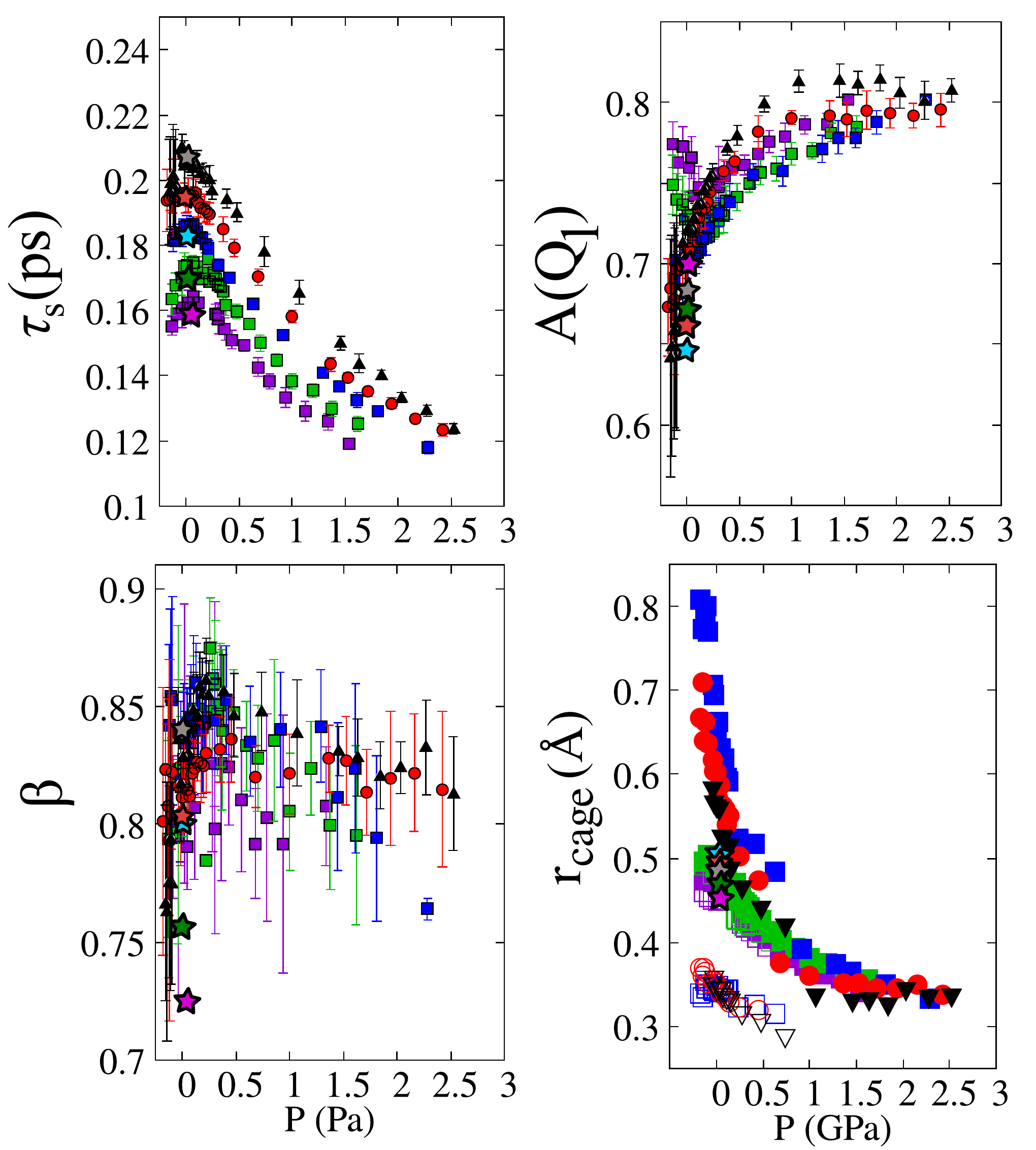}
    \caption{Results of fits of the self-intermediate scattering function $F(Q_1,t)$ using Eq.~\ref{eq:exp_fit}:  (a) short-time relaxation time $\tau_\mathrm{s}$, (b) amplitude $A(Q_1)$ and (c) Kohlrausch exponent $\beta$ as a function of pressure $P$ for $T =$ {\color{black}220~K (purple), 240~K (green),} 260~K (blue squares), 280~K (red circles) and 300 K (black triangles).  Stars symbols shown have been obtained at a fixed $Q=2.25$ Å$^{-1}$ in Ref.~\onlinecite{de2016mode}. In panel (d) we present cage radius $r_\mathrm{cage}$ as a function of pressure $P$ using Eq~\ref{eq:cage_rad}, computed for $T =$ {\color{black}220~K (purple), 240~K (green),} 260~K (blue squares), 280~K (red circles) and 300 K (black triangles). Note that filled and empty symbols correspond to $Q_1$ and $Q_2$, respectively. }
    \label{fig:Fk_260_280}
\end{figure}

The correlations in both space and time \textcolor{black}{are} characterized by the self-intermediate scattering function, $F(Q,t)$ (Eq.~\ref{eq:F}). Fitting $F(Q,t)$ with Eq. \ref{eq:exp_fit}, the simulated values can be analyzed in terms of four parameters: the $\alpha$-relaxation time $\tau_{\alpha}$, the short-time relaxation time $\tau_s$, the non-ergodicity parameter $A(k)$, and the Kohlrausch exponent $\beta$ as functions of pressure. The results \textcolor{black}{at the wavevector $Q_1$, the location of the first maximum in $S(Q)$ (see Section~\ref{sec:Sq})}, are displayed in Figs.~\ref{fig:tau_alpha} and \ref{fig:Fk_260_280} \textcolor{black}{and listed in Tables~\ref{table:fitting220} to \ref{table:fitting300}}.

\begin{figure}[t]
    \centering
    \includegraphics[width=0.95\linewidth]{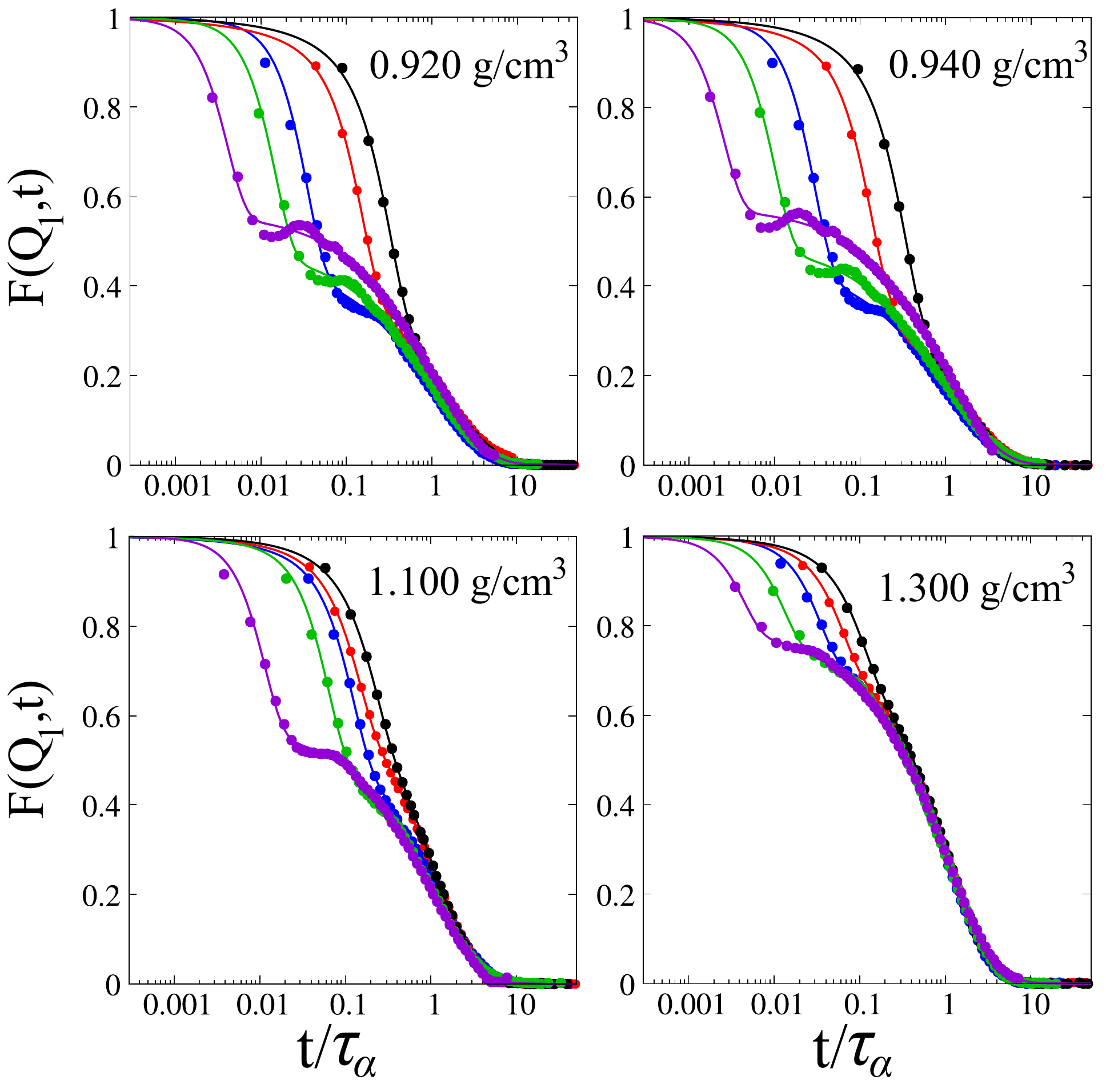}
    \caption{Illustration of the time-temperature superposition principle~\cite{de2016mode} for $T =$ {\color{black} 220~K (purple), 240~K (green),} 260~K (blue), 280~K (red) and 300 K (black).}
    \label{fig:TTSP}
\end{figure}

It was already noted in previous simulations\cite{de2016mode} that $\tau_\alpha$ in TIP4P/2005 water is anomalous: in contrast to normal liquids, it decreases with increasing pressure. Our results are in excellent agreement with the previous simulations, and, by reaching more than 30\% higher densities, they reveal the existence of a minimum in $\tau_\alpha${\color{black}, located at similar pressures at all temperatures.} The present simulations thus confirm the existence of a new anomaly in water, a minimum in its relaxation time along isotherms, previously proposed based on the analysis of experimental data~\cite{eichler2025shear}. The experimental values for $\tau_\alpha$, measured at 295 K, are shown by a solid red curve in Fig.~\ref{fig:tau_alpha}. As noted above for viscosities, the experimental curve is better reproduced by the simulations at 280 K.

From the fitting parameters that describe $F(Q,t)$, one can define $r_\mathrm{cage}$, the radius of the cage in which the molecules rattle at the early stage of structural relaxation (Eq.~\ref{eq:cage_rad}). Figure \ref{fig:Fk_260_280} (d) presents the dependence of $r_\mathrm{cage}$ with pressure at the two first maxima in $S(Q)$, $Q_1$ and $Q_2$ (Fig.~\ref{fig:Sq_vs_dens}). The apparent disagreement with Ref.~\citenum{de2016mode} arises from the difference in the chosen $Q$, which was fixed at $2.25$ Å$^{-1}$ in that work. At $Q_2$, $r_\mathrm{cage}$ decreases sharply with applied pressure, witnessing the strong rearrangement occuring in the second coordination shell. At $Q_1$, $r_\mathrm{cage}$ first decreases slowly with increasing pressure, and suddenly jumps up around 0.7 GPa. This coincides with the disappearance of $Q_2$ at the structural crossover, when the second shell has fully collapsed over the first. At higher pressure, $r_\mathrm{cage}$ resumes a slow decrease. This structural metric provides a direct link between the features observed in $S(q)$ and the changing local environment governing molecular dynamics.

\begin{figure*}[tb]
    \centering
    \includegraphics[width=0.95\linewidth]{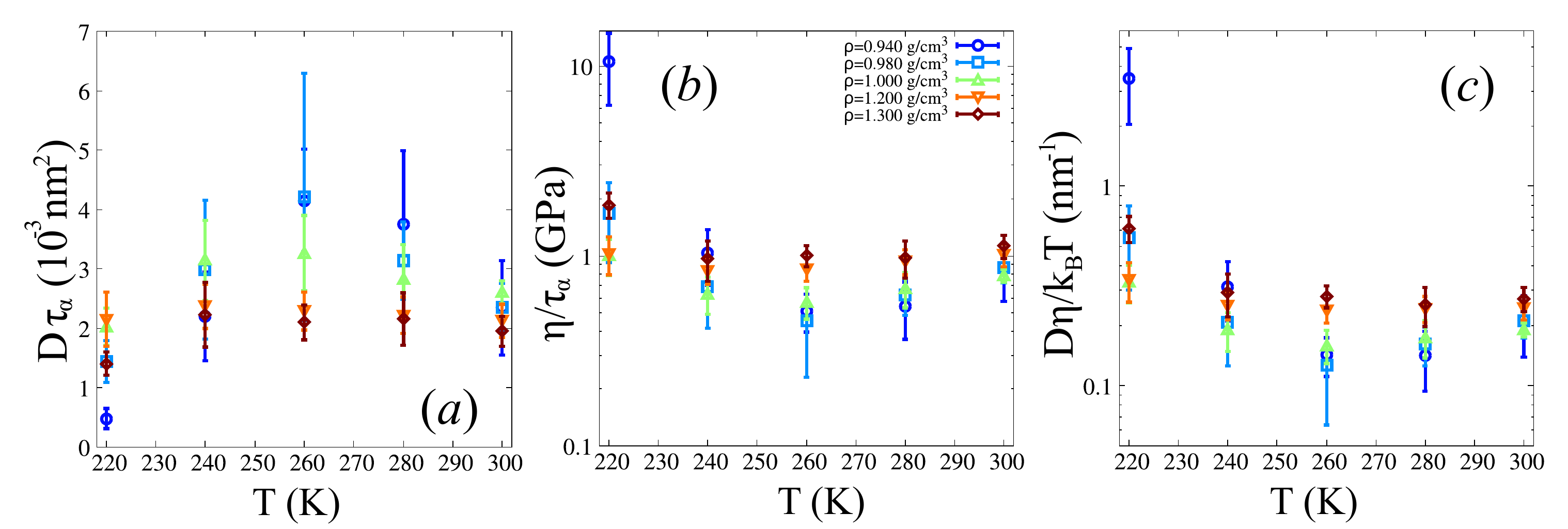}
    \caption{{\color{black} Test of the coupling between dynamic quantities along several isochores: $D \tau_\alpha$ (a), $\eta/\tau_\alpha$ (b), and $D\eta/T$ (c) as a function of temperature $T$.  
    Data from Ref.~\onlinecite{montero2018viscosity} were interpolated with a parabolic fit with weights taken from the uncertainty to obtain $D$ and $\eta$ at the same density at which $\tau_\alpha$ was computed in the present work.}}
    \label{fig:Dtaueta_Temp}
\end{figure*}

\begin{figure}[!]
    \centering
    \includegraphics[width=0.92\linewidth]{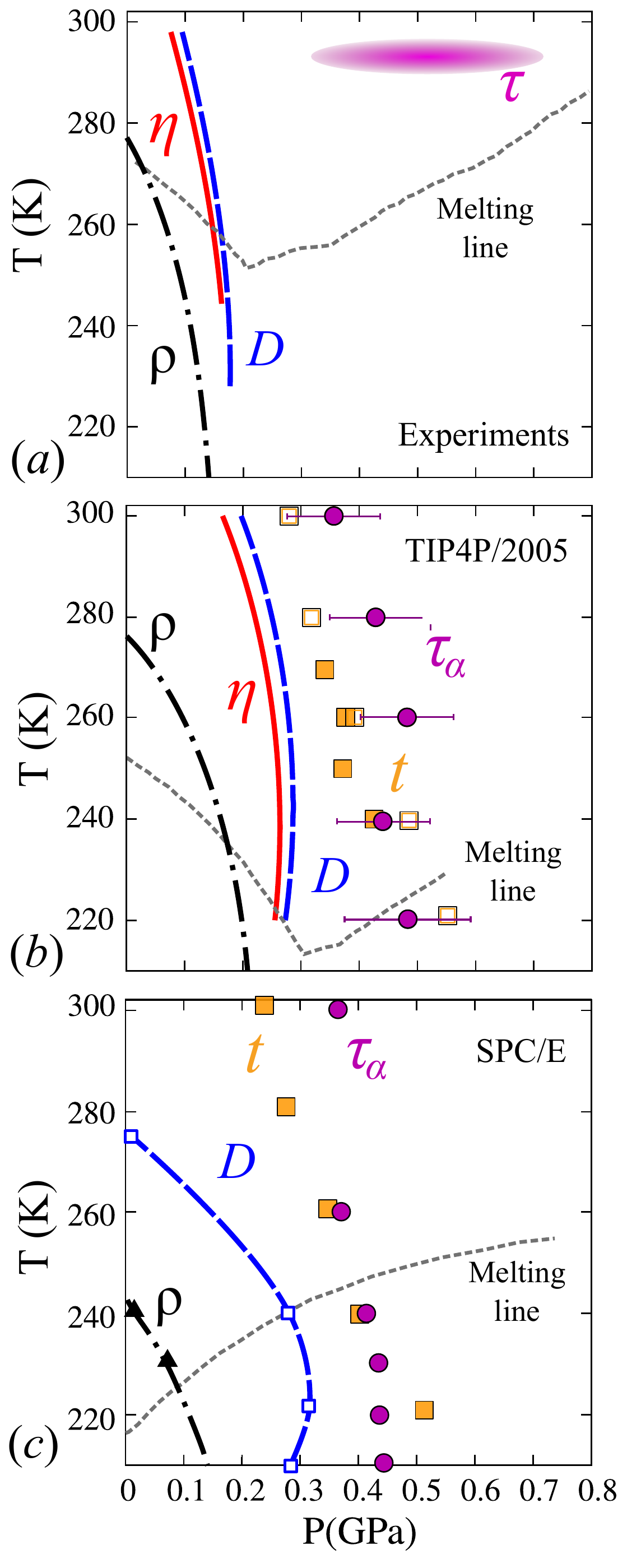}
    \caption{Lines of anomalies in liquid water for experiments (a) {\color{black} and simulations for TIP4P/2005  (b) and  SPC/E (c)}. Extrema along isotherms are shown: density maximum (dash-dotted black), minimum of $\eta$ (solid red), maximum of $D$ (long dashed blue), minimum of $t$ (orange squares), and minimum of $\tau_\alpha$ (purple ellipse and circles). The short dashed curves show the respective melting lines. {\color{black}For real water and TIP4P/2005, the lines of $\rho$, $\eta$, and $D$ extrema were obtained from fits with a two-state model\cite{singh2017pressure,montero2018viscosity}. For SPC/E, we took data form the literature to draw the melting line\cite{vega2005can} and the lines of anomalies for $\rho$\cite{harrington1997equation}, $D$\cite{starr1999slow}, $t$\cite{errington2001relationship}, and $\tau_\alpha$\cite{starr1999dynamics}.} }
    \label{fig:lines}
\end{figure}

We can comment further on the shape of $F(Q,t)$. First, according to mode-coupling theory \textcolor{black}{(MCT)\cite{Gotze_relaxation_1992}}, it should obey the time-temperature superposition principle (TTSP): at long times, 
$F(Q,t)$ at various temperatures for a given isochore should collapse on a master curve if plotted against the rescaled time $t/\tau_{\alpha}$. This was shown to be the case for TIP4P/2005 at low densities~\cite{de2016mode}. We confirm TTSP also holds at higher densities (see Fig.~\ref{fig:TTSP}). 

At 300 K, $F(Q,t)$ exhibits a single-step, nearly exponential decay for all densities, indicative of simple liquid-like behavior where high thermal energy dominates the relaxation process. This result aligns perfectly with previous simulation studies of water models at ambient temperatures \cite{de2016mode}. At 280 K, an incipient separation of timescales begins to emerge, and at {\color{black}lower temperatures} the relaxation clearly shows two steps. The intermediate plateau is usually interpreted as the manifestation of the trapping of a water molecule inside the cage formed by its neighbours, before the cage relaxes and allows for the final $\alpha$ relaxation. As already noticed in Ref.~\onlinecite{de2016mode}, this plateau is more developped at lower pressures: the liquid at low density, with better defined first and second coordination shell (see Section~\ref{sec:g}) is more structurally arrested.

\section{Discussion\label{sec:discussion}}

Simulations of shear and bulk viscosities with the TIP4P/2005 model show a good agreement with the pressure variation measured in experiments at room temperature~\cite{eichler2025shear}, and provide a prediction for its temperature dependence. We expect that, at lower temperatures, a pressure minimum of bulk viscosity should become more apparent, and that the ratio of bulk to shear should exhibit a more pronounced decrease at high pressure.

Our simulations also give a more detailed insight than available experiments about the structure and structural relaxation at extreme pressures. Based on the experimental results of Ref.~\onlinecite{eichler2025shear}, it was proposed that the structural relaxation time along the room temperature isotherm goes through a minimum around 0.5 GPa. However, this was only indirect, based on analysis of viscosity and sound \textcolor{black}{attenuation} data with a visco-elastic model. Here in simulations, we can access directly the structural relaxation time. The results fully confirm the existence of a minimum in $\tau_\alpha(P)$, thus further validating the analysis performed in Ref.~\onlinecite{eichler2025shear}, and predict that it shifts to higher pressures upon cooling.

\textcolor{black}{We can also discuss the degree of coupling between various dynamic quantities. It is often assumed that the self-diffusion coefficient $D$ and $1/\tau_\alpha$ have a very similar behavior. This idea is supported by MCT, which in its idealized version predicts a power-law divergence at low temperature of both quantities with the same exponent\cite{Gotze_relaxation_1992}, and consequently a constant product $D \tau_\alpha$ along isochores. With SPC/E, it was found that $D \tau_\alpha$ is nearly constant at high temperature along each isochore, but increases slightly at the lowest temperature\cite{Gallo_slow_1996,starr1999dynamics}. Figure~\ref{fig:Dtaueta_Temp} (a) shows $D \tau_\alpha$ along selected isochores for TIP4P/2005. In the temperature range considered, the two quantities vary by several orders of magnitude along an isochore, but their product is remarkably less sensitive to temperature, making the error bars more apparent in Fig.~\ref{fig:Dtaueta_Temp}. At high density, we find a nearly constant product. At lower densities, $D\tau_\alpha$ starts increasing on cooling, similarly to SPC/E. However, upon further cooling, it starts decreasing again, and for the lowest density even reaches a value lower than at $300\,\mathrm{K}$. It has been already observed in a study of atomic binary mixtures with softened repulsive interactions\cite{Shi_relaxation_2013} that $D\tau_\alpha$ along isochores may deviate from a constant value on cooling. But, in that case, $D\tau_\alpha$ went through a temperature minimum instead of a maximum, and the degree of deviation depended on density and on the specific interaction potential. We also see here that the choice of water potential affects the qualitative variation of $D\tau_\alpha$. Another relation that is often assumed in the literature is the proportionality between $\eta$ and $\tau_\alpha$, although deviations have also been observed\cite{Shi_relaxation_2013}. Figure~\ref{fig:Dtaueta_Temp} (b) shows $\eta/\tau_\alpha$ along selected isochores for TIP4P/2005. Between 240 and 300 K, $\eta/\tau_\alpha$ is nearly constant at high density; a minimum develops as the density decreases. At 220 K, $\eta/\tau_\alpha$ increases, with an amplitude that varies non-monotonically with increasing density: extreme at the lowest density, it first decreases before increasing mildly at the highest densities.  This shows that, although the relations $D \tau_\alpha  =\mathrm{cst}$ or $\eta/ \tau_\alpha  =\mathrm{cst}$ are useful for deducing a correct order of magnitude for $D$ and $\eta$ from $\tau_\alpha$ or vice-versa, they should be used with caution. For completeness, we also display in Fig.~\ref{fig:Dtaueta_Temp} (c) $D \eta/T$, already reported in our previous work\cite{montero2018viscosity}. The Sutherland-Stokes-Einstein relation, which predicts $D \eta/T = \mathrm{cst}$, is violated at low temperature. {\color{black} Experimentally, the pressure range in which viscosity data is available for highly supercooled water limits the test of the Stokes-Einstein-Sutherland relation to 0-250 MPa. The violation gets less pronounced with increasing pressure ~\cite{dehaoui2015viscosity,mussa2023viscosity}, and is in good agreement with TIP4P/2005 simulations ~\cite{Dueby_decoupling_2019,Dubey_understanding_2019}. In TIP4P/2005 simulations, which span a broader pressure range than experiments, the degree of violation depends non-monotonically on density (Fig.~\ref{fig:Dtaueta_Temp}  (c)), similar to $\eta/\tau$ (Fig. ~\ref{fig:Dtaueta_Temp} (b)).} This violation upon cooling has been attributed to the increasing contribution of jump motion to the diffusion of water molecules\cite{Dueby_decoupling_2019,Dubey_understanding_2019}. Jump motion may also play a role in the decoupling of $\tau_\alpha$ from $D$ and $\eta$ in deeply supercooled water; this point deserves further investigation.}

We now turn to the speculation formulated in Ref.~\onlinecite{eichler2025shear}, that the minimum in $\tau_\alpha$ could be related to a structural change. Fig.~\ref{fig:lines} shows the loci of all anomalies of \textcolor{black}{experimental data (a), TIP4P/2005 (b), and SPC/E (c)} models in the pressure-temperature plane. \textcolor{black}{This representation is motivated by the work of Errington and Debenedetti\cite{errington2001relationship}, who introduced it to reveal the cascade of anomalies present in the phase diagram of SPC/E. Agarwal~\textit{et al.} obtained a similar nested pattern for TIP4P/2005\cite{agarwal2011thermodynamic}. However, in these works, the $\tau_\alpha$ anomaly was not considered. We display the line of $\tau_\alpha$ minima in Fig.~\ref{fig:lines}, using our results for TIP4P/2005, and the data from Ref.~\onlinecite{starr1999dynamics} for SPC/E.} From the inside \textcolor{black}{(low $T$ and $P$)} to the outside \textcolor{black}{(high $T$ and $P$)}, the full sequence \textcolor{black}{of anomalies} now reads: density maximum, diffusivity maximum, viscosity minimum, translational order parameter minimum, and relaxation time minimum. \textcolor{black}{This representation confirms the imperfect coupling between $\eta$, $D$, and $\tau_\alpha$ discussed above: their extrema along isotherms occur on separate lines. Overall, the agreement with experiments is better with TIP4P/2005 than with SPC/E, especially for the density and diffusivity maxima, and for the melting line; this stems from the fact that SPC/E describes a liquid that is understructured compared to TIP4P/2005 or real water\cite{starr1999slow}. The translational order parameter minimum has not yet been identified in experiments, but simulations for both SPC/E and TIP4P/2005 show that it occurs close to the line of $\tau_\alpha$ minima. This suggests that, indeed, as proposed in Ref.~\onlinecite{eichler2025shear}, the $\tau_\alpha$ anomaly may have its origin in a major change in the translational order of water. In 1948, Hall used a two-state picture for water to describe sound attenuation and interpreted $\tau_\alpha$ as the typical time for interconversion between the two local structures\cite{hall1948origin}. $\tau_\alpha$ could go through a minimum when the structural anomaly in translational order kinetically facilitates interconversion.}

\textcolor{black}{Our results, computed over a wide range of densities and temperatures, contribute to complete the map of water anomalies. Among all properties, the relaxation time $\tau_{\alpha}$ plays a prominent role, as it defines the broadest anomalous region in the phase diagram (in which $\tau_{\alpha}$ decreases with applied pressure, contrary to normal liquids). This calls for experiments at other temperatures to determine the locus of $\tau_\alpha$ minima.}

\begin{acknowledgments}
C.V. acknowledges fundings IHRC22/00002 and  PID2022-140407NB-C21 funded by MCIN/AEI /10.13039/501100011033 and by  FEDER, UE.

This work is dedicated to Carlos Vega, with whom most of the authors have collaborated in the last 14 years. Working with Carlos has been a pleasure and a luxury not only for being a recognised researcher but because of his enthusiasm and humbleness. His wise scientific and personal advices have been really helpful for many of his collaborators and students who still learn from him.

\end{acknowledgments}


\section*{Data Availability Statement}

The data that support the findings of this study are available from the corresponding author upon reasonable request. 

\appendix

\section{Values of $\kappa$, $\eta$, and $D$}

Table~\ref{table: Diffusion} gives the values $\kappa$, $\eta$, and $D$ simulated along the isotherms at the three highest temperatures.

\begin{table}[hb]
\caption{\label{table: Diffusion} {\color{black}Values at the three highest studied temperatures} $T$ (K) and densities $\rho$ (g cm$^{-3}$) of the pressure $P$ (GPa), the bulk viscosity $\kappa$ (mPa s), the shear viscosity $\eta$ (mPa s), and the self-diffusion coefficient $D$ ($10^{-9}\,\mathrm{m^{2}\, s^{-1}}$)  corrected with Eq.~\ref{eq:diffus2}.}
\begin{ruledtabular}
\begin{tabular}{c|ccccc}
T & $\rho$ & $P$ & $\kappa$  & $\eta$ & $D$\\
\hline
& 0.920 & -0.133 & - & 1348 $\pm$ 281 & 0.00497 $\pm$ 0.00067 \\
& 0.960 & -0.038 & - & 164  $\pm$ 46  & 0.0207 $\pm$ 0.0028  \\
& 1.000 & 0.047 &  - & 36.8 $\pm$ 5.7 & 0.0576 $\pm$ 0.0020 \\
& 1.050 & 0.113 &  - & 20.3 $\pm$ 2.0 & 0.0976 $\pm$ 0.0033  \\
220 & 1.130 & 0.280 &  - & 13.8 $\pm$ 1.3 & 0.1455 $\pm$ 0.0060  \\
& 1.150 & 0.367 &  - & 14.4 $\pm$ 1.4 & 0.1490 $\pm$ 0.0061 \\
& 1.200 & 0.550 &  - & 14.8 $\pm$ 1.6 & 0.1353 $\pm$ 0.0056 \\
& 1.250 & 0.790 &  - & 22.6 $\pm$ 1.7 & 0.1044 $\pm$ 0.0073 \\
& 1.275 & 0.940 &  - & 33.7 $\pm$ 5.0 & 0.0715 $\pm$ 0.0050 \\
& 1.325 & 1.340 &  - & 78.2 $\pm$ 12 &  0.0392 $\pm$ 0.0027 \\
\hline
&0.920 & -0.137 & - & 45.3 $\pm$ 9.0  &  0.065 $\pm$ 0.011   \\
&0.960 & -0.042 & - & 15.9 $\pm$1.5  &   0.144 $\pm$ 0.024  \\
&1.000 & 0.008 &  -&  7.94 $\pm$0.56 &   0.255 $\pm$ 0.010  \\
&1.050 & 0.105 & - & 5.40  $\pm$0.41 &  0.342 $\pm$ 0.014   \\
240 &1.130 & 0.308 & - & 4.24 $\pm$ 0.15 &  0.453 $\pm$ 0.022 \\
&1.150 & 0.371 & - & 4.40 $\pm$ 0.24 &  0.437 $\pm$ 0.021 \\
&1.200 & 0.592 & - & 4.83 $\pm$ 0.28  & 0.402 $\pm$ 0.019 \\
&1.275 & 1.001 & - & 8.21 $\pm$ 0.23 & 0.2504 $\pm$ 0.0076 \\
&1.325 & 1.379 & - & 12.81 $\pm$ 0.24 &  0.1742 $\pm$ 0.0053 \\
\hline
    & 0.997 & 0.008  & 7.30 $\pm$ 0.22 & 2.430 $\pm$ 0.060&0.288 $\pm$ 0.001   \\  
    & 1.050 & 0.106  & 6.60 $\pm$ 0.25 & 2.103 $\pm$ 0.067& 0.282 $\pm$ 0.001 \\
    & 1.100 & 0.237  & 6.15 $\pm$ 0.27 & 1.974 $\pm$ 0.069 & 0.306 $\pm$ 0.001\\ 
    & 1.175 & 0.518  & 6.56 $\pm$ 0.27& 2.145 $\pm$ 0.062&0.285 $\pm$ 0.001 \\
    & 1.260 & 0.911  & 8.78 $\pm$ 0.34& 3.053 $\pm$ 0.086& 0.240 $\pm$ 0.001 \\
260 & 1.300 & 1.289  & 11.10 $\pm$ 0.60 & 4.053 $\pm$ 0.132&0.223  $\pm$ 0.001\\
    & 1.320 & 1.451  & 11.40 $\pm$ 1.90  & 4.766 $\pm$ 0.167& 0.167 $\pm$ 0.001 \\
    & 1.330 & 1.533  & 14.20 $\pm$ 0.62 & 5.249 $\pm$ 0.198& 0.160 $\pm$ 0.001 \\
    & 1.340 & 1.616  & 14.06 $\pm$ 0.73& 5.521 $\pm$ 0.227 &0.169 $\pm$ 0.001 \\
    & 1.350 & 1.714  & 14.51 $\pm$ 0.66 & 6.572 $\pm$ 0.218& 0.166 $\pm$ 0.001 \\
\hline
    & 0.997 &  0.002  &3.74 $\pm$ 0.16 &1.302  $\pm$  0.028  &0.551 $\pm$  0.003 \\
    & 1.050	&  0.112   &3.65 $\pm$ 0.12 &1.227  $\pm$  0.023 &0.545 $\pm$ 0.005 \\
    & 1.100	&  0.250   &3.49 $\pm$ 0.12 &1.181  $\pm$  0.023 &0.604 $\pm$ 0.005 \\
    & 1.175	&  0.569   &4.18 $\pm$ 0.15 &1.403  $\pm$  0.029 &0.66 $\pm$ 0.003\\
    & 1.260	&  1.003   &4.97 $\pm$ 0.20 &1.850   $\pm$  0.040  &0.394 $\pm$ 0.003\\
280 & 1.300 &  1.368   &6.24 $\pm$ 0.30 & 2.364  $\pm$  0.052 &0.413 $\pm$ 0.001 \\
    & 1.320	&  1.527   &7.08 $\pm$ 0.27 & 2.563  $\pm$  0.074 &0.263 $\pm$ 0.002 \\
    & 1.330	&  1.622   &7.08  $\pm$ 0.32 & 2.783 $\pm$  0.087 &0.311 $\pm$ 0.003 \\
    & 1.340	&  1.718   &8.22 $\pm$ 0.29 & 3.185  $\pm$  0.096 &0.325 $\pm$ 0.003 \\
    & 1.350	&  1.828   &8.19 $\pm$ 0.35 & 3.330  $\pm$  0.069  &0.237 $\pm$  0.003\\
\hline
    &0.997 &  0.007  & 2.00 $\pm$ 0.06 &0.796  $\pm$ 0.030 & 0.796 $\pm$ 0.004 \\
    &1.050  & 0.133  & 2.25 $\pm$ 0.05&0.799   $\pm$ 0.028 & 0.824 $\pm$ 0.008 \\
    &1.100  & 0.273  & 2.24 $\pm$ 0.05& 0.809  $\pm$ 0.037 & 0.538 $\pm$ 0.001 \\ 
    &1.175  & 0.484  & 2.54 $\pm$ 0.04 &0.931  $\pm$ 0.044 &0.719 $\pm$ 0.007 \\
    &1.260  & 1.069  & 4.06 $\pm$ 0.44&1.307   $\pm$ 0.092 &0.581 $\pm$ 0.002 \\
    &1.300  & 1.461  & 3.79 $\pm$ 0.06 &1.597  $\pm$ 0.119& 0.411 $\pm$ 0.001 \\
300    &1.320  & 1.633  & 5.53 $\pm$ 0.76 &1.852  $\pm$ 0.078 &0.363 $\pm$ 0.001 \\
    & 1.330 & 1.768  & 3.66 $\pm$ 0.68 & 1.961 $\pm$ 0.012 &0.255 $\pm$ 0.001 \\
    & 1.340 & 1.858  & 5.32 $\pm$ 0.06 & 2.013 $\pm$ 0.011 &0.237 $\pm$ 0.001 \\ 
    & 1.350 & 1.950  & 5.94 $\pm$ 0.59 & 2.254 $\pm$ 0.008 &0.216 $\pm$ 0.001 \\ 
    & 1.370 & 2.141  & 6.05 $\pm$ 0.44 & 2.546 $\pm$ 0.011 &0.192 $\pm$ 0.001 \\
    & 1.400 & 2.443  & 8.17 $\pm$ 0.90 & 3.281 $\pm$ 0.039 &0.152 $\pm$ 0.001 \\ 
    & 1.420 & 2.764  &10.11 $\pm$ 0.86 & 3.872 $\pm$ 0.046 & - \\
\end{tabular}
\end{ruledtabular}
\end{table}

\section{Structural relaxation}

In what follows we report \textcolor{black}{five tables summarizing the fitting parameters of Eq.~\ref{eq:exp_fit} at T=220 K (Table~\ref{table:fitting220}), 240 K ((Table~\ref{table:fitting240}),} T=260 K (Table~\ref{table:fitting260}), 280 K ((Table~\ref{table:fitting280}), and T=300K (Table~\ref{table:fitting300}).

\begin{table}[t]
\caption{\label{table:fitting220}Parameters obtained at $T=220$ K: density $\rho$ (g cm$^{-3}$), pressure $P$ (GPa), amplitude $A$, short-time relaxation time $\tau_s$ (ps), long-time relaxation time $\tau_c$ (ps), and Kohlrausch exponent $\beta$.}
\begin{ruledtabular}
\begin{tabular}{ccccccc}
$\rho$ & $P$ & $Q_1$ & $\tau_\alpha$ & $\tau_s$ & $A$ & $\beta$ \\
\hline
0.920 & -0.133 & 1.85 & 75.04 & 0.155 & 0.77 & 0.82 \\
0.940 & -0.086 & 1.86 & 45.55 & 0.159 & 0.76 & 0.80 \\
0.960 & -0.038 & 1.86 & 54.63 & 0.160 & 0.77 & 0.79 \\
0.980 & 0.015 & 1.90& 40.67 & 0.162 & 0.76 & 0.83 \\
1.000 & 0.047 & 1.93 & 35.93 & 0.160 & 0.76 & 0.79 \\
1.025 & 0.070 & 2.02 & 25.91 & 0.164 & 0.74 & 0.83 \\
1.050 & 0.113 & 2.04 & 23.70 & 0.162 & 0.75 & 0.81 \\
1.130 & 0.280 & 2.14 & 19.29 & 0.158 & 0.75 & 0.83 \\
1.135 & 0.309 & 2.14 & 20.53 & 0.157 & 0.76 & 0.79 \\
1.140 & 0.302 & 2.17 & 18.68 & 0.158 & 0.74 & 0.85 \\
1.150 & 0.367 & 2.18 & 15.70 & 0.154 & 0.75 & 0.82 \\
1.175 & 0.433 & 2.18 & 17.09 & 0.151 & 0.76 & 0.82 \\
1.200 & 0.550 & 2.23 & 15.73 & 0.149 & 0.76 & 0.81 \\
1.225 & 0.678 & 2.26 & 20.08 & 0.142 & 0.76 & 0.79 \\
1.250 & 0.790 & 2.28 & 22.08 & 0.138 & 0.77 & 0.80 \\
1.275 & 0.940 & 2.32 & 25.36 & 0.133 & 0.78 & 0.79 \\
1.300 & 1.126 & 2.35 & 26.27 & 0.128 & 0.78 & 0.77 \\
1.325 & 1.340 & 2.37 & 32.64 & 0.125 & 0.78 & 0.80 \\
1.350 & 1.537 & 2.39 & 61.02 & 0.119 & 0.80 & 0.71 \\
\end{tabular}
\end{ruledtabular}
\end{table}

\begin{table}[t]
\caption{\label{table:fitting240}Parameters obtained in the fit of the self-intermediate scattering function $F(Q_1,t)$ with Eq.~\ref{eq:exp_fit} at $T=240$ K: density $\rho$ (g cm$^{-3}$), pressure $P$ (GPa), wavevector $Q_1$ (\AA), structural relaxation time $\tau_\alpha$ (ps), short-time relaxation time $\tau_s$ (ps), amplitude $A$, and Kohlrausch exponent $\beta$.}
\begin{ruledtabular}
\begin{tabular}{ccccccc}
$\rho$ & $P$ & $Q_1$ & $\tau_\alpha$ & $\tau_s$ & $A$ & $\beta$ \\
\hline
0.920 & -0.137 & 1.88 & 32.44 & 0.163 & 0.75 & 0.79 \\
0.940 & -0.100 & 1.89 & 22.31 & 0.168 & 0.74 & 0.80 \\
0.960 & -0.042 & 1.90 & 22.12 & 0.170 & 0.74 & 0.82 \\
0.980 & -0.013 & 1.95 & 15.60 & 0.173 & 0.73 & 0.82 \\
1.000 & 0.008 & 1.97 & 12.66 & 0.174 & 0.73 & 0.83 \\
1.025 & 0.062 & 2.01 & 10.33 & 0.175 & 0.72 & 0.84 \\
1.050 & 0.105 & 2.06 & 9.07 & 0.173 & 0.72 & 0.85 \\
1.070 & 0.147 & 2.14 & 9.52 & 0.170 & 0.73 & 0.85 \\
1.080 & 0.159 & 2.12 & 6.63 & 0.172 & 0.72 & 0.86 \\
1.090 & 0.208 & 2.12 & 4.48 & 0.176 & 0.72 & 0.84 \\
1.100 & 0.227 & 2.14 & 4.80 & 0.169 & 0.73 & 0.78 \\
1.110 & 0.258 & 2.15 & 6.58 & 0.171 & 0.72 & 0.87 \\
1.115 & 0.255 & 2.14 & 5.52 & 0.171 & 0.73 & 0.85 \\
1.120 & 0.280 & 2.16 & 5.16 & 0.170 & 0.73 & 0.86 \\
1.125 & 0.306 & 2.20 & 4.47 & 0.168 & 0.73 & 0.86 \\
1.130 & 0.308 & 2.19 & 5.69 & 0.168 & 0.73 & 0.86 \\
1.135 & 0.325 & 2.17 & 6.09 & 0.167 & 0.73 & 0.85 \\
1.140 & 0.350 & 2.20 & 6.01 & 0.166 & 0.73 & 0.85 \\
1.150 & 0.371 & 2.19 & 6.77 & 0.162 & 0.74 & 0.84 \\
1.175 & 0.477 & 2.22 & 5.74 & 0.160 & 0.74 & 0.85 \\
1.200 & 0.592 & 2.24 & 5.93 & 0.156 & 0.75 & 0.83 \\
1.225 & 0.700 & 2.27 & 7.01 & 0.150 & 0.76 & 0.83 \\
1.250 & 0.853 & 2.31 & 8.14 & 0.145 & 0.76 & 0.84 \\
1.275 & 1.001 & 2.33 & 9.81 & 0.138 & 0.77 & 0.81 \\
1.300 & 1.194 & 2.36 & 10.54 & 0.135 & 0.77 & 0.82 \\
1.325 & 1.379 & 2.37 & 13.57 & 0.130 & 0.78 & 0.80 \\
1.350 & 1.619 & 2.40 & 19.84 & 0.125 & 0.78 & 0.80 \\

\end{tabular}
\end{ruledtabular}
\end{table}

\begin{table}[t]
\caption{\label{table:fitting260}Parameters obtained in the fit of the self-intermediate scattering function $F(Q_1,t)$ with Eq.~\ref{eq:exp_fit} at $T=260$ K: density $\rho$ (g cm$^{-3}$), pressure $P$ (GPa), wavevector $Q_1$ (\AA), structural relaxation time $\tau_\alpha$ (ps), short-time relaxation time $\tau_s$ (ps), amplitude $A$, and Kohlrausch exponent $\beta$.}
\begin{ruledtabular}
\begin{tabular}{ccccccc}
$\rho$ & $P$ & $Q_1$ & $\tau_\alpha$ & $\tau_s$ & $A$ & $\beta$ \\
\hline
 0.930 &-0.130 & 1.94 & 8.96 & 0.181 & 0.69 & 0.84\\
 0.940 &-0.123 & 1.90 & 9.05 & 0.182 & 0.70 &0.85 \\
 0.950 &-0.103 & 1.96 & 7.09 & 0.183 & 0.69 &0.85 \\
 0.980 &-0.044 & 2.02 & 6.91 & 0.184 & 0.70 &0.84 \\
 0.990 &-0.015 & 2.00 & 5.73 & 0.186 & 0.71 &0.81 \\
 0.997 & 0.008 & 2.03 & 4.76 & 0.188 & 0.70 & 0.83 \\
 1.010 & 0.017 & 2.04 & 4.29 & 0.186 & 0.69 &0.84 \\
 1.020 & 0.043 & 2.06 & 3.64 & 0.186 & 0.71 &0.83 \\
 1.030 & 0.059 & 2.08 & 3.26 & 0.186 & 0.71 &0.84 \\
 1.040 & 0.091 & 2.09 & 3.58 & 0.183 & 0.70 &0.85 \\
 1.050 & 0.109 & 2.09 & 3.13 & 0.180 & 0.71 & 0.85 \\
 1.060 & 0.117 & 2.13 & 3.34 & 0.182 & 0.71 &0.86 \\
 1.100 & 0.237 & 2.18 & 3.11 & 0.178 & 0.73 & 0.84 \\
 1.150 & 0.408 & 2.18 & 2.93 & 0.170 & 0.74 &0.85 \\
 1.175 & 0.518 & 2.23 & 2.93 & 0.172 & 0.74 & 0.85 \\
 1.200 & 0.619 & 2.27 & 2.82 & 0.162 & 0.75 &0.83 \\
 1.250 & 0.929 & 2.32 & 3.60 & 0.152 & 0.76 &0.84 \\
 1.260 & 0.911 & 2.33 & 3.60 & 0.150 & 0.75 & 0.84 \\
 1.300 & 1.289 & 2.37 & 4.20 & 0.141 & 0.77 & 0.84 \\
 1.320 & 1.451 & 2.38 & 5.34 & 0.136 & 0.78 & 0.81 \\
 1.340 & 1.616 & 2.40 & 6.34 & 0.132 & 0.78 & 0.82 \\
 1.360 & 1.714 & 2.41 & 6.63 & 0.129 & 0.79 & 0.79 \\
 1.400 & 2.299 & 2.45 & 10.97 & 0.117 & 0.80 & 0.76 \\
\end{tabular}
\end{ruledtabular}
\end{table}

\begin{table}[t]
\caption{\label{table:fitting280} Parameters obtained in the fit of the self-intermediate scattering function $F(Q_1,t)$ with Eq.~\ref{eq:exp_fit} at $T=280$ K: density $\rho$ (g cm$^{-3}$), pressure $P$ (GPa), wavevector $Q_1$ (\AA), structural relaxation time $\tau_\alpha$ (ps), short-time relaxation time $\tau_s$ (ps), amplitude $A$, and Kohlrausch exponent $\beta$.}
\begin{ruledtabular}
\begin{tabular}{ccccccc}
$\rho$ & $P$ & $Q_1$ & $\tau_\alpha$ & $\tau_s$ & $A$ & $\beta$ \\
\hline
 0.910 & -0.176 & 2.07 & 3.67 & 0.194 & 0.67 & 0.80 \\
 0.920 & -0.153 & 1.97 & 3.63 & 0.194 & 0.68 & 0.82 \\
 0.930 & -0.138 & 2.05 & 2.98 & 0.195 & 0.67 & 0.81 \\
 0.940 & -0.121 & 2.01 & 3.18 & 0.194 & 0.68 & 0.79 \\
 0.950 & -0.109 & 2.04 & 2.78 & 0.196 & 0.70 & 0.82 \\
 0.980 & -0.048 & 2.08 & 2.30 & 0.195 & 0.70 & 0.82 \\
 0.990 & -0.021 & 2.09 & 1.97 & 0.196 & 0.70 & 0.80 \\
 0.997 &  0.002 & 2.09 & 1.95 & 0.19 & 0.70 & 0.81 \\
 1.010 &  0.021 & 2.04 & 1.76 & 0.196 & 0.71 & 0.82 \\
 1.020 &  0.032 & 2.06 & 1.75 & 0.195 & 0.72 & 0.81 \\
 1.030 &  0.055 & 2.08 & 1.63 & 0.195 & 0.72 & 0.81 \\
 1.040 &  0.076 & 2.09 & 1.46 & 0.196 & 0.73 & 0.82 \\
 1.050 &  0.112 & 2.18 & 1.75 & 0.19 & 0.73 & 0.82 \\
 1.060 &  0.139 & 2.14 & 1.77 & 0.192 & 0.73 & 0.83 \\ 
 1.100 &  0.250 & 2.20 & 1.40 & 0.18 & 0.75 & 0.83 \\
 1.175 &  0.569 & 2.25 & 1.45 & 0.18 & 0.77 & 0.84 \\
 1.200 &  0.692 & 2.29 & 1.58 & 0.170 & 0.78 & 0.82 \\
 1.250 &  0.996 & 2.33 & 1.75 & 0.158 & 0.79 & 0.82 \\
 1.260 & 1.003 & 2.34 & 1.75 & 0.16 & 0.79 & 0.82 \\
 1.300 & 1.368 & 2.38 & 2.37 & 0.14 & 0.79 & 0.83 \\
 1.320 & 1.527 & 2.40 & 2.58 & 0.14 & 0.79 & 0.83 \\
 1.340 & 1.718 & 2.41 & 3.19 & 0.14 & 0.80 & 0.81 \\
 1.360 & 1.828 & 2.42 & 3.42 & 0.13 & 0.79 & 0.82 \\
 1.380 & 2.122 & 2.40 & 4.25 & 0.13 & 0.79 & 0.82 \\
 1.400 & 2.317 & 2.45 & 4.84 & 0.12 & 0.80 & 0.81 \\
\end{tabular}
\end{ruledtabular}
\end{table}

\begin{table}[t]
\caption{\label{table:fitting300} Parameters obtained in the fit of the self-intermediate scattering function $F(Q_1,t)$ with Eq.~\ref{eq:exp_fit} at $T=300$ K: density $\rho$ (g cm$^{-3}$), pressure $P$ (GPa), wavevector $Q_1$ (\AA), structural relaxation time $\tau_\alpha$ (ps), short-time relaxation time $\tau_s$ (ps), amplitude $A$, and Kohlrausch exponent $\beta$.}
\begin{ruledtabular}
\begin{tabular}{ccccccc}
$\rho$ & $P$ & $Q_1$ & $\tau_\alpha$ & $\tau_s$ & $A$ & $\beta$ \\
\hline
 0.980 & -0.024 & 2.15 & 0.97 & 0.210 & 0.71 & 0.84 \\
 0.990 & -0.009 & 2.18 & 1.04 & 0.205 & 0.70 & 0.82 \\
 0.997 &  0.007 & 2.18 & 1.05 & 0.207 & 0.72 & 0.83 \\
 1.010 &  0.032 & 2.17 & 1.03 & 0.205 & 0.72 & 0.84 \\
 1.020 &  0.049 & 2.24 & 0.96 & 0.205 & 0.72 & 0.83 \\
 1.030 &  0.068 & 2.27 & 1.01 & 0.204 & 0.73 & 0.84 \\
 1.040 &  0.090 & 2.20 & 0.97 & 0.205 & 0.74 & 0.85 \\
 1.050 &  0.133 & 2.18 & 0.96 & 0.20 & 0.74 & 0.85 \\
 1.060 &  0.169 & 2.22 & 0.92 & 0.202 & 0.75 & 0.86 \\ 
 1.100 & 0.273 & 2.22 & 0.91 & 0.197 & 0.77 & 0.86 \\
 1.175 & 0.484 & 2.27 & 1.03 & 0.193 & 0.78 & 0.85 \\
 1.200 & 0.744 & 2.32 & 1.02 & 0.178 & 0.80 & 0.85 \\
 1.250 & 1.059 & 2.35 & 1.13 & 0.166 & 0.81 & 0.84 \\
 1.260 & 1.069 & 2.35 & 1.13 & 0.168 & 0.81 & 0.84 \\
 1.300 & 1.461 & 2.38 & 1.41 & 0.150 & 0.81 & 0.83 \\
 1.320 & 1.633 & 2.40 & 1.61 & 0.144 & 0.81 & 0.83 \\
 1.340 & 1.858 & 2.41 & 1.68 & 0.140 & 0.82 & 0.82 \\
 1.360 & 1.950 & 2.35 & 2.14 & 0.133 & 0.81 & 0.82 \\
 1.380 & 2.141 & 2.44 & 2.47 & 0.129 & 0.80 & 0.83 \\
 1.400 & 2.443 & 2.39 & 2.93 & 0.124 & 0.81 & 0.81 \\
\end{tabular}
\end{ruledtabular}
\end{table}

\nocite{*}
\section*{Bibliography}
\bibliography{bibliography}

\end{document}